\documentclass[fleqn,usenatbib]{mnras}

\usepackage{newtxtext,newtxmath}

\usepackage[T1]{fontenc}

\DeclareRobustCommand{\VAN}[3]{#2}
\let\VANthebibliography\thebibliography
\def\thebibliography{\DeclareRobustCommand{\VAN}[3]{##3}\VANthebibliography}

\usepackage{graphicx}	
\usepackage{amsmath}	
\usepackage[flushleft,referable]{threeparttablex}
\usepackage{longtable}
\usepackage{natbib}
\usepackage{enumitem}
\usepackage{pdflscape} 

\newcommand{\SSFR}{$\Sigma_{\rm SFR}$}
\newcommand{\sSSFR}{$\Sigma_{\rm sSFR}$}

\newcommand{\Vmax}{$V_{\rm{max}}$}
\newcommand{\Vcir}{$v_{\rm{circ}}$}
\newcommand{\Vesc}{$v_{\rm{esc}}$}
\newcommand{\RE}{$R_{\rm{E}}$}
\newcommand{\HB}{$\rm{H}\beta$}
\newcommand{\OIII}{[\ion{O}{III}]}
\newcommand{\HA}{$\rm{H}\alpha$}
\newcommand{\NII}{[\ion{N}{II}]}
\newcommand{\Mste}{$M_{\star}$}
\newcommand{\Mdot}{$\dot{M}_{\rm{out}}$}
\newcommand{\Mload}{$\eta_{m}$}
\newcommand{\Eload}{$\eta_{E}$}
\newcommand{\Pload}{$\eta_{p}$}

\title[MOSDEF Survey: Ionised Outflows]{The MOSDEF Survey: Properties of Warm Ionised Outflows at $z$ = 1.4 -- 3.8}

\author[A. Weldon et al.]{
Andrew Weldon,$^{1}$\thanks{E-mail: aweld004@ucr.edu}
Naveen A. Reddy,$^{1}$
Alison L. Coil,$^{2}$
Alice E. Shapley,$^{3}$
Brian Siana,$^{1}$
\newauthor
Mariska Kriek,$^{4}$
Bahram Mobasher,$^{1}$
Zhiyuan Song,$^{1}$
Michael A. Wozniak,$^{1}$
\\
$^{1}$Department of Physics and Astronomy, University of California, Riverside, 900 University Avenue, Riverside, CA 92521, USA\\
$^{2}$Center for Astrophysics and Space Sciences, Department of Physics, University of California, San Diego, 9500 Gilman Dr., La Jolla, CA 92093-0424, USA\\
$^{3}$Physics \& Astronomy Department, University of California, Los Angeles, 430 Portola Plaza, Los Angeles, CA 90095, USA\\
$^{4}$Leiden Observatory, Leiden University, PO Box 9513, NL-2300 RA Leiden, The Netherlands\\
}

\date{Accepted XXX. Received YYY; in original form ZZZ}

\pubyear{2024}

\begin{document}
\label{firstpage}
\pagerange{\pageref{firstpage}--\pageref{lastpage}}
\maketitle

\begin{abstract}
We use the large spectroscopic data set of the MOSFIRE Deep Evolution Field survey to investigate the kinematics and energetics of ionised gas outflows. Using a sample of 598 star-forming galaxies at redshift 1.4 < $z$ < 3.8, we decompose \HA\ and \OIII\ emission lines into narrow and broad components, finding significant detections of broad components in 10\% of the sample. The ionised outflow velocity from individual galaxies appears independent of galaxy properties, such as stellar mass, star-formation rate (SFR), and star-formation-rate surface density (\SSFR). Adopting a simple outflow model, we estimate the mass-, energy- and momentum-loading factors of the ionised outflows, finding modest values with averages of 0.33, 0.04, and 0.22, respectively. The larger momentum- than energy-loading factors, for the adopted physical parameters, imply that these ionised outflows are primarily momentum-driven. We further find a marginal correlation (2.5$\sigma$) between the mass-loading factor and stellar mass in agreement with predictions by simulations, scaling as \Mload\ $\propto M_{\star}^{-0.45}$. This shallow scaling relation is consistent with these ionised outflows being driven by a combination of mechanical energy generated by supernovae explosions and radiation pressure acting on dusty material. In a majority of galaxies, the outflowing material does not appear to have sufficient velocity to escape the gravitational potential of their host, likely recycling back at later times. Together, these results suggest that the ionised outflows traced by nebular emission lines are negligible, with the bulk of mass and energy carried out in other gaseous phases.
\end{abstract}




\section{Introduction}

Large-scale galactic outflows have long been recognised as a key process in galaxy evolution. Early theoretical models of galaxy formation required outflows of mass and energy to prevent gas from excessively cooling, causing overly efficient star formation and an overproduction of stellar mass \citep{White78, Dekel86, White91}. Modern models of galaxy evolution and simulations rely on intense, galactic-scale outflows to suppress star formation and reproduce observed properties of galaxies, such as the galaxy mass function and the sizes of galactic disks and bulges \citep[e.g.,][]{Guedes11, Dave11}. Beyond suppressing star-formation, outflows modulate the metallicity within galaxies, enriching the circumgalactic medium (CGM) and possibly the intergalactic medium (IGM) with metals \cite[e.g.,][]{Tremonti04, Dalcanton07, Finlator08}. At late-times, $z \lesssim$ 1, recycled gas from past outflows may fuel a significant fraction of star-formation in galaxies \citep{Oppenheimer10, Henriques13, Alcazar17}. Outflows also appear to be an important factor in the creation of low-column-density channels in the ISM, allowing for the escape of ionising photons \citep[e.g.,][]{Gnedin08, Leitet13, Ma16, Reddy16, Gazagnes18, Reddy22}.

Despite advances in numerical modelling, a complete description of galactic-scale outflows and their impacts on galaxy evolution is challenging for simulations due to the wide scale range of outflows. Large-scale cosmological hydrodynamical suites such as EAGLE \citep{Schaye15} and Illustris TNG \citep{Pillepich18} do not resolve the small scales required to capture relevant feedback processes that generate outflows. Instead, these models employ sub-grid processes for feedback, which vary between simulations. On the other hand, high-resolution “local patch” simulations capture the interaction between stellar feedback and the multi-phase ISM but lack the size needed to track the long-term evolution of outflows \cite[e.g.,][]{Girichidis16, Girichidis18, Li17, Kim18, Kim20}. In either case, observational constraints on how the properties of outflows scale with the stellar mass, star-formation rate, and other properties of their host across galaxy populations are essential.

Quantifying the impact of outflows on galaxy evolution based on observational data has been an active area of research for well over a decade. Multi-wavelength observations have shown that galactic outflows have a multi-phase structure, with outflows detected in hot X-ray ($\sim$10$^{6-7}$ K) emitting gas, from emission and absorption lines tracing warm ($\sim$10$^{4}$ K) and cool ($\sim$10$^{3}$ K) gas, and down to cold ($\lesssim$100 K) molecular and dust outflows in radio observations \citep[see reviews by][]{Heckman17, Rupke18, Veilleux20}. In typical star-forming galaxies, outflows are theorised to be driven by energy injected into the ISM by supernovae; radiation pressure acting on cool, dusty material; cosmic rays; or a combination of these mechanisms \citep{Ipavich75, Chevalier85, Murray05, Murray11}. Galactic-scale outflows are then expected to play a major role in galaxy evolution at $z \sim$ 1 -- 3, during the peak of the cosmic star-formation history \citep["Cosmic Noon";][]{Madau14}, when feedback from star formation and AGN was maximised. 

At these redshifts, outflows are a common feature of star-forming galaxies, with blueshifted rest-UV interstellar absorption lines tracing cool, neutral/low-ionisation outflows \citep[e.g.,][]{Shapley03, Steidel10, Weldon22} and broad components of rest-optical emission lines tracing warm, ionised outflows \citep[hereafter ionised outflows; e.g.,][]{Genzel11, Genzel14, Newman12, Freeman19, Concas22}. Observations of cool outflows have found that outflow velocity increases with several galactic properties, such as stellar mass, star-formation rate (SFR), and star-formation-rate surface density (\SSFR) \citep[e.g.,][]{Rubin10, Steidel10, Martin12, Rubin14, Chisholm15, Weldon22}. Absorption lines are sensitive to gas along the entire line of sight, including both high- and low-density gas from current and past outflows, thus they likely trace material ejected over long timescales. Estimates of mass outflow rates from absorption lines will then depend on the metallicity of the outflowing gas, the outflow geometry, and the absorption contribution from the ISM and faint satellite galaxies, which are often unconstrained by absorption line observations \citep[e.g.,][]{Weiner09, Heckman15}. On the other hand, the underlying broad components of strong rest-optical emission lines likely trace denser outflowing gas near the launching points of the outflows, providing a snapshot of current outflow activity. The mass outflow rates traced by emission lines depend on similar unconstrained (e.g., outflow geometry) and other properties that may be estimated from emission lines (e.g., the electron density of the outflowing gas). However, due to the difficulty of detecting faint broad components in typical $z \gtrsim$ 1 galaxies, studies are primarily limited to local galaxies, small samples, gravitationally lensed galaxies, or high S/N composite spectra to infer average outflow properties \citep{Newman12, Davies19, Freeman19, Swinbank19, Schreiber19, Concas22, Chu22a, Chu22b}. 

The mass-loading factor (\Mload) is a key characteristic of outflows, which represents the amount of mass they remove normalised by the galaxy’s star-formation rate and, in star-forming galaxies, is thought of as a proxy for outflow efficiency. Simple analytical arguments and numerical simulations predict an anti-correlation between the mass-loading factor and outflow velocity or stellar mass of galaxies, which scales steeper if the outflows are energy-driven and shallower if they are momentum-driven \citep[e.g.,][]{Murray05, Oppenheimer08, Muratov15}. Observations of ionised outflows in local galaxies find such trends, suggesting that outflows are more efficient at removing material from the shallower potential wells of lower-mass galaxies \citep[e.g.,][]{Heckman15, Chisholm17, McQuinn19, Marasco23}. In addition to the mass-loading factor, outflows are characterised by the amount of energy and momentum they carry with respect to the amounts generated by supernovae and stellar winds. However, at high redshifts, there are few constraints on the mass-, energy-, and momentum-loading factors of outflows in typical star-forming galaxies.

In this paper, we expand upon the work of \cite{Freeman19} to characterise the properties of ionised gas outflows from individual star-forming galaxies at $z \sim$ 2 using the complete MOSFIRE Deep Evolution Field survey \citep[MOSDEF;][]{MOSDEF}. The MOSDEF survey obtained rest-optical spectra for $\sim$1500 high-redshift galaxies, most with multiple emission lines for which we can investigate outflows in several emission lines (i.e, \HB, \OIII, \HA, and \NII) from a large sample of galaxies. Our goals for this study are to constrain the kinematics and loading factors of ionised gas outflows and to explore how these properties are related to galactic properties, such as stellar mass, SFR, and \SSFR. The outline of this paper is as follows. In Section \ref{sec:Data}, we introduce the sample, measurements of galaxy properties, and the methodology of fitting galaxy spectra to characterise the presence of outflows. Section \ref{sec:results}, presents our main results on the correlations between measured galaxy properties, ionised outflow velocities, and mass-loading factors. We discuss the physical context behind these results in Section \ref{sec:discussion} and summarise our conclusions in Section \ref{sec:conclusions}. Throughout this paper, we adopt a standard cosmology with $\Omega_{\Lambda}$ = 0.7, $\Omega_{M}$ = 0.3, and $\rm{H}_{0}$ = 70 km $\rm{s}^{-1} \rm{ Mpc}^{-1}$. All wavelengths are presented in the vacuum frame.

\section{Data and Measurements}
\label{sec:Data}

\subsection{MOSDEF Survey}
\label{sec:survey}

Galaxies analysed in this paper were drawn from the MOSDEF survey, which targeted $\approx$1500 $H$-band selected galaxies and AGNs at redshifts 1.4 $\leq$ $z$ $\leq$ 3.8 in the CANDELS fields \citep{Grogin11, Koekemoer11}. The survey obtained moderate-resolution ($R$ $\sim$ 3000--3600) rest-optical spectra using the Multi-Object Spectrometer for Infra-Red Exploration \citep[MOSFIRE;][]{McLean12} on the Keck I telescope. Galaxies were targeted for spectroscopy based on pre-existing spectroscopic, grism, or photometric redshifts that placed them in three redshift ranges ($z$ = 1.37 -- 1.70, $z$ = 2.09 -- 2.61, and $z$ = 2.95 -- 3.80). This selection optimised the coverage of several strong rest-frame optical emission lines ([\ion{O}{II}]$\lambda\lambda$3727,3730, \HB, [\ion{O}{III}]$\lambda\lambda$4960,5008, \HA, [\ion{N}{II}]$\lambda\lambda$6550,6585, and [\ion{S}{II}]$\lambda\lambda$6717,6732) that lie in the YJHK transmission windows. The final MOSDEF sample spans ranges of star-formation rate (1 < SFR < 200 M$_{\odot}$ yr$^{-1}$) and stellar mass (10$^{9}$ < \Mste\ < 10$^{11}$ M$_{\odot}$) typical for galaxies at $z \sim$ 1.4 -- 3.8, with the majority of galaxies having detections of multiple rest-frame optical emission lines. For full details regarding the MOSDEF survey (targeting, data reduction, and sample properties), we refer readers to \cite{MOSDEF}.

Emission-line fluxes were measured by simultaneously fitting a line with the best-fit SED model for the continuum and a Gaussian function for the line \citep[see][for a complete description of the SED modelling]{Reddy22}. For multiple lines that lie in close proximity, multiple Gaussians were fit, such as the [\ion{O}{II}] doublet and \HA\ and the \NII\ doublet, which were fitted with two and three Gaussians, respectively. Systemic redshifts were derived from the strongest emission line, usually \HA\ or [\ion{O}{III}]$\lambda$5008, and were used to fit the other rest-frame optical nebular emission lines. Further details on emission-line measurements and slit loss corrections are given in \cite{MOSDEF} and \cite{Reddy15}. 

Galaxy sizes and inclinations were estimated from the effective radius (\RE), within which half the total light of the galaxy is contained, and the axis ratio ($b$/$a$), respectively, measured by \cite{vanderWel14}\footnote{\url{https://users.ugent.be/~avdrwel/research.html}} using GALFIT \citep{Peng10} on HST/F160W images from the CANDELS survey. 

\subsection{Sample Selection}
\label{sec:selection}

In our analysis, we search for underlying, broad rest-optical emission components from star-forming galaxies, tracing ionised outflowing gas. The parent MOSDEF sample contains 878 galaxies with \OIII\ and 759 with \HA\ detections. Several criteria are applied to the parent MOSDEF sample to create a sample conducive for measuring broad emission. First, 94 \OIII\ and 104 \HA\ detections from active galactic nuclei (AGNs) identified by IR colors, X-ray emission, and/or the \NII/\HA\ line ratio were removed \citep{Coil15, Azadi17, Azadi18, Leung19}. Next, we removed: detections with S/N < 10; detections that were affected by bright skylines; and detections that lay close to the edge of their spectra, reducing the sample to 431 galaxies with \OIII\ and 514 with \HA\ detections. Finally, galaxies may have broad line components simply from rotation and velocity dispersion. As such, to isolate outflowing broad emission, we limit the narrow emission to FWHM < 275 km s$^{-1}$ (see Section \ref{sec:fitting}) and remove detections with FWHM > 275 km s$^{-1}$ from a single Gaussian fit. These criteria result in a final sample of 598 galaxies (hereafter the “MOSDEF-ionised” sample), of which 391 (435) have \OIII\ (\HA) detections. There are 228 galaxies with both an \OIII\ and \HA\ detection.

\subsection{Galaxy Properties}
\label{sec:gal_props}

In this study, we investigate the properties of ionised outflows against several global galaxy properties (e.g., stellar mass, SFR, star-formation-rate surface density). Stellar masses (\Mste), SFRs, ages, and color excesses were derived from spectral energy distribution (SED) modelling. Here, we briefly describe the models used and refer readers to \cite{Reddy15} for more details. The models were created adopting a \citet[hereafter BC03]{Bruzual03}  stellar population synthesis model, \cite{Chabrier03} initial mass function, constant star formation histories (SFH), Small Magellanic Cloud (SMC) attenuation curve \citep{Fitzpatrick90, Gordon03}, and sub-solar metallicity ($Z_{\ast} = 0.28 Z_{\odot}$). A lower age limit of 50 Myr was imposed, based on the typical dynamical timescale of $z \sim$ 2 galaxies \citep{Reddy12}. The combination of the steeper SMC attenuation curve, which has been found to best reproduce the dust obscurations of typical star-forming galaxies at $z \sim$ 2 based on far-infrared data \citep{Reddy18a}, and sub-solar metallicity provide self-consistent SFRs with those derived using other methods \citep{Reddy18b, Theios19}. The best-fit stellar population parameters and their errors were obtained by perturbing the photometry, refitting the models, and taking the median and dispersion in the resulting parameters, respectively.

Blueshifted interstellar absorption lines tracing cool, neutral/low-ionisation gas outflows are ubiquitous in $z \gtrsim$ 2 star-forming galaxies \citep[e.g.,][]{Shapley03, Steidel10}. Observations suggest that the velocity of cool outflows increases with the SFR and \SSFR\ of a galaxy \citep[e.g.,][]{Steidel10, Martin12, Rubin14, Weldon22}. As the broad components of rest-optical emission lines likely trace denser outflowing gas near the launching points of the outflows, they provide a snapshot of current outflow activity, thus their velocity may also scale with SFR and \SSFR. We calculate \HA\ SFRs (SFR[\HA]) from \HA\ and \HB\ flux measurements corrected for dust using the Balmer decrement. Following the methodology presented in \cite{Reddy15}, \HA\ luminosities are corrected for attenuation assuming a \cite{Cardelli89} Galactic extinction curve and converted to SFRs using the conversion factor from \cite{Reddy18b}, $\rm{3.236\times10^{-42}}$ $M_{\odot}$ yr$^{-1}$ ergs$^{-1}$ s, for a BC03 stellar population synthesis model and sub-solar metallicity adopted for the SED fitting. SFR[\HA] is calculated for objects with significant detections (S/N > 3) of \HA\ and \HB. As discussed in previous studies, there is a general agreement between SFR[SED] and SFR[\HA] for MOSDEF galaxies \citep[e.g.,][]{Reddy15, Shivaei16, Azadi18, Reddy22}.

When an ionised outflow is detected, SFR[\HA] should be derived from the narrow flux component tracing gas within the galaxy. The \HA\ SFRs above are then overestimates as they use \HA\ and \HB\ fluxes measured from single-Gaussian fits. As the stellar continuum is unlikely affected by emission from an outflow, SFR[SED] should be insensitive to the presence of outflowing gas. On the other hand, when detected in \HA, we can correct SFR[\HA] by multiplying it by the narrow-to-single \HA\ flux ratio. For these reasons, in our analysis, we have chosen to focus on SFR[SED] when discussing outflows detected in \OIII\ and SFR[\HA] when detected in \HA. 

Along with the star-formation rate, the mechanisms that drive outflows may be enhanced in regions of compact star formation. We define the star-formation-rate surface density as \SSFR\ = SFR/(2$\pi R_{\rm{E}}^{2}$). Additionally, at a given \SSFR, outflows may be more effectively launched from a shallow galaxy potential (i.e., low stellar mass) relative to a deep potential \citep{Pucha22, Reddy22}. To examine the frequency of galaxies with observed outflows on both \SSFR\ and the galaxy potential, we define the specific star-formation-rate surface density as \sSSFR\ = SFR/(2$\pi R_{\rm{E}}^{2} M_{\star}$)

\subsection{Searching for Broad Emission Lines}
\label{sec:fitting}

\subsubsection{Fitting Individual Galaxies}
\label{sec:fit_indiv}

We search for ionised gas outflows by decomposing \HB, [\ion{O}{III}]$\lambda\lambda$4960,5008 and \HA, [\ion{N}{II}]$\lambda\lambda$6550,6585 into narrow Gaussian components, tracing virial motions within the galaxy, and broad Gaussian components tracing the ionised outflowing gas. In the most general case, simultaneously fitting narrow and broad Gaussians to each set of three emission lines would require 19 free parameters. Motivated by previous studies \citep[e.g.,][]{Genzel11, Genzel14, Newman12}, we adopt the following assumptions: (1) the narrow components of each line share the same FWHM (FWHM$_{\rm{na}}$) and redshift, (2) the broad components of each line share the same FWHM (FWHM$_{\rm{br}}$) and velocity offset from the narrow component ($\Delta v_{\rm{br}}$), and (3) the [\ion{O}{III}]$\lambda$5008/[\ion{O}{III}]$\lambda$4960 and [\ion{N}{II}]$\lambda$6585/[\ion{N}{II}]$\lambda$6550 flux ratios are 2.98 and 2.93, respectively \citep{Osterbrock89}. Therefore, each fit has nine free parameters; five shared by each line (FWHM$_{\rm{na}}$, FWHM$_{\rm{br}}$, $\Delta v_{\rm{br}}$, narrow component redshift, and constant background) and four controlling the narrow and broad component amplitudes ($A_{\rm{na}}$ and $A_{\rm{br}}$). 

For each set of lines, we perform two preliminary fits and one final fit. The first preliminary fit uses a linear continuum and single Gaussians to fit the emission lines using \texttt{curve fit}, a non-linear least squares fitting routine from the \texttt{scipy.optimize} subpackage. We use this fit to subtract off the linear continuum and normalise the spectra by the peak of the brightest line for each set of lines (\OIII\ or \HA). Next, using the normalised spectra, we fit each emission line with a narrow and broad Gaussian with \texttt{curve fit}. The resulting values of the second fit are used as initial values for the final fit, which is done using \texttt{emcee}, a Python Markov chain Monte Carlo (MCMC) Ensemble sampler \citep{emcee}. We take the median values of the resulting posterior probability distributions for all the model parameters. The errors on the parameters are estimated using the 16th and 84th percentiles.

\begin{figure*}
  \includegraphics[width=\linewidth, keepaspectratio]{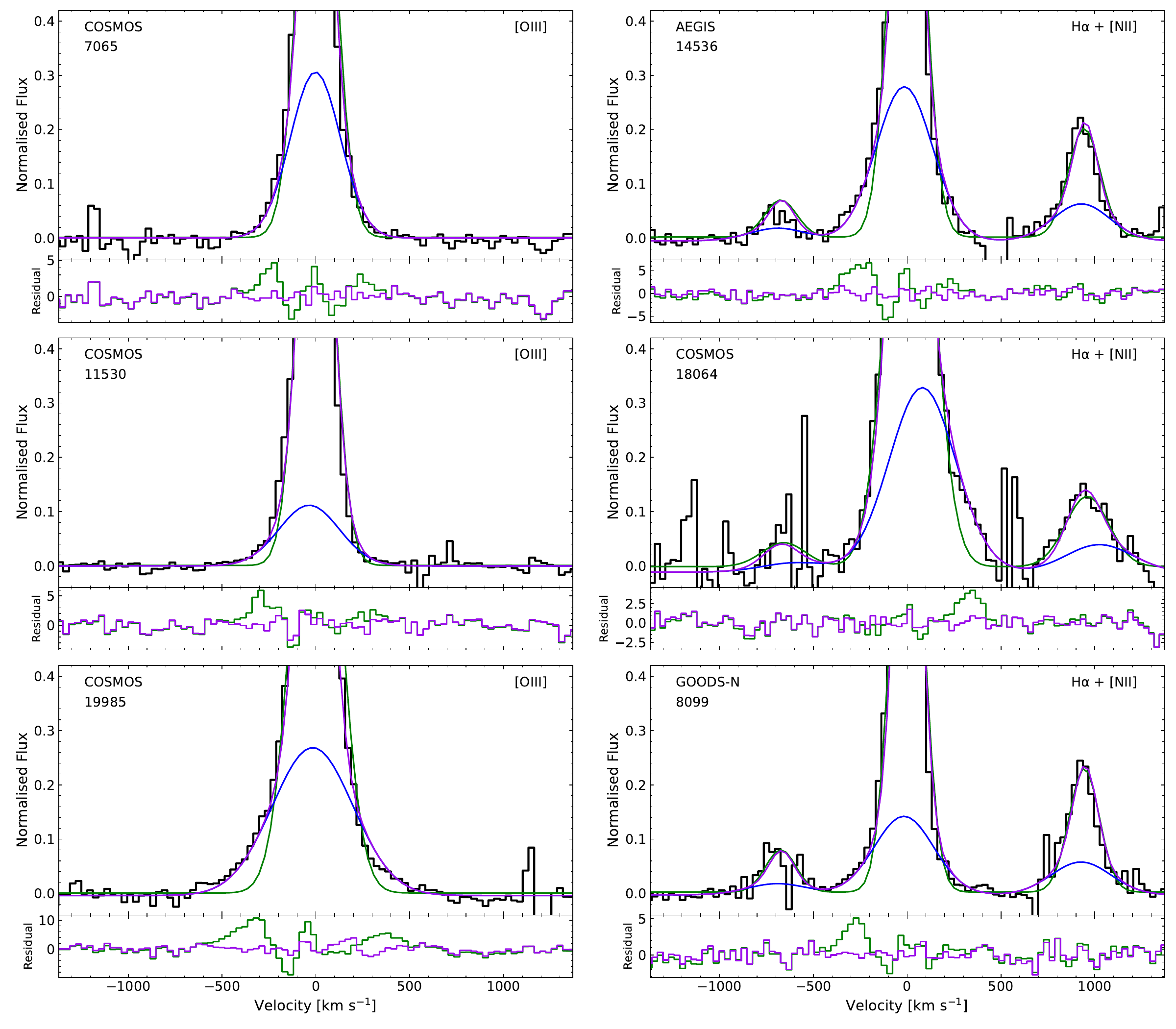}
  \vspace{-0.45cm}
  \caption{Examples of fits for individual spectra with large $\Delta$BIC values. The field and 3D-HST v4.1 catalog ID is given in the upper left. Each line is normalised such that the strongest line peak is unity. The single Gaussian fit is shown in green, the overall fit for the narrow+broad components is shown in purple, and the broad component is shown in blue. The bottom plot in each panel shows the residuals after subtracting the best-fit single (green) and narrow+broad fit (purple) from the spectrum. The "wave" pattern, an underestimation flux at the peak and overestimation in the wing(s), in the residuals shows that a single Gaussian does not fit the observed line profiles well.}
  \label{fig:example}
\end{figure*}

In order to properly study ionised outflows, we must be certain that the broad components trace a kinematically distinct feature from the rotation of the host galaxy, thus we place physically motivated restrictions on the free parameters. In particular, we restricted the FWHM$_{\rm{na}}$ to values between 80 and 275 km s$^{-1}$ and FWHM$_{\rm{br}}$ to values between 300 and 800 km s$^{-1}$. The lower limit on FWHM$_{\rm{na}}$ is the average skyline FWHM of the MOSDEF-ionised galaxies, while a majority of galaxies (90\%) exhibit an FWHM < 275 km s$^{-1}$  when fitting \OIII\ or \HA\ with a single Gaussian component. The additional 25 km s$^{-1}$  separation between FWHM$_{\rm{na}}$ and FWHM$_{\rm{br}}$ helps ensure that the broad component is not an artefact from a better fit to the narrow emission by using two Gaussian components. Typical values of FWHM$_{\rm{br}}$ for ionised outflows from star-forming galaxies are 300--600 km s$^{-1}$  \citep{Genzel11, Newman12, Wood15}. The centroids of the narrow and broad components are limited to within $\pm$100 km s$^{-1}$ of their initial values, as found by similar studies \citep{Newman12, Wood15, Davies19, Concas22}. 

To determine whether a broad component is detected, we evaluate the improvement over a single Gaussian fit using the Bayesian Information Criterion \citep[BIC;][]{Schwarz78} and the amplitude of the broad component. Following a similar procedure as for the double Gaussian fit, we fit each set of lines with single Gaussians: the spectra are first normalised using an initial fit, then refitted, the results of which serve as the initial values for a MCMC fitting process \footnote{For the single fits there are five free parameters: constant background, redshift, FWHM, and two amplitudes ($A_{\rm{H\beta}}$ and $A_{\rm{[\ion{O}{III}]}}$ or  $A_{\rm{H\alpha}}$ and $A_{\rm{[\ion{N}{II}]}}$).}. The BIC is defined as: 
\begin{equation}
    \text{BIC} = \chi^{2} + k \text{ln}\left(n \right),
\end{equation}
where $\chi^{2}$ is the chi squared of the fit, $k$ is the number of parameters used in the fit, and $n$ is the number of points used in the fit. Following similar studies, we adopt $\Delta$BIC = BIC$_{\rm{single}}$ -- BIC$_{\rm{double}}$ > 10 as "very strong" evidence against a single Gaussian fit \citep[e.g.,][]{Swinbank19, Avery21, Concas22}. Additionally, to ensure that the broad component is not an artefact, we require that the broad component amplitude of \OIII\ or \HA\ is robustly measured (A$_{\rm{br}}$ $-$ 3$\sigma_{\rm{A_{br}}}$ > 0) as evidence for the detection of a broad component. Figure \ref{fig:example} presents fits for six galaxies that show strong evidence for a broad component. 

\subsubsection{Detectability of Broad Components}
\label{sec:mock}

\begin{figure}
    \includegraphics[width=\columnwidth, keepaspectratio]{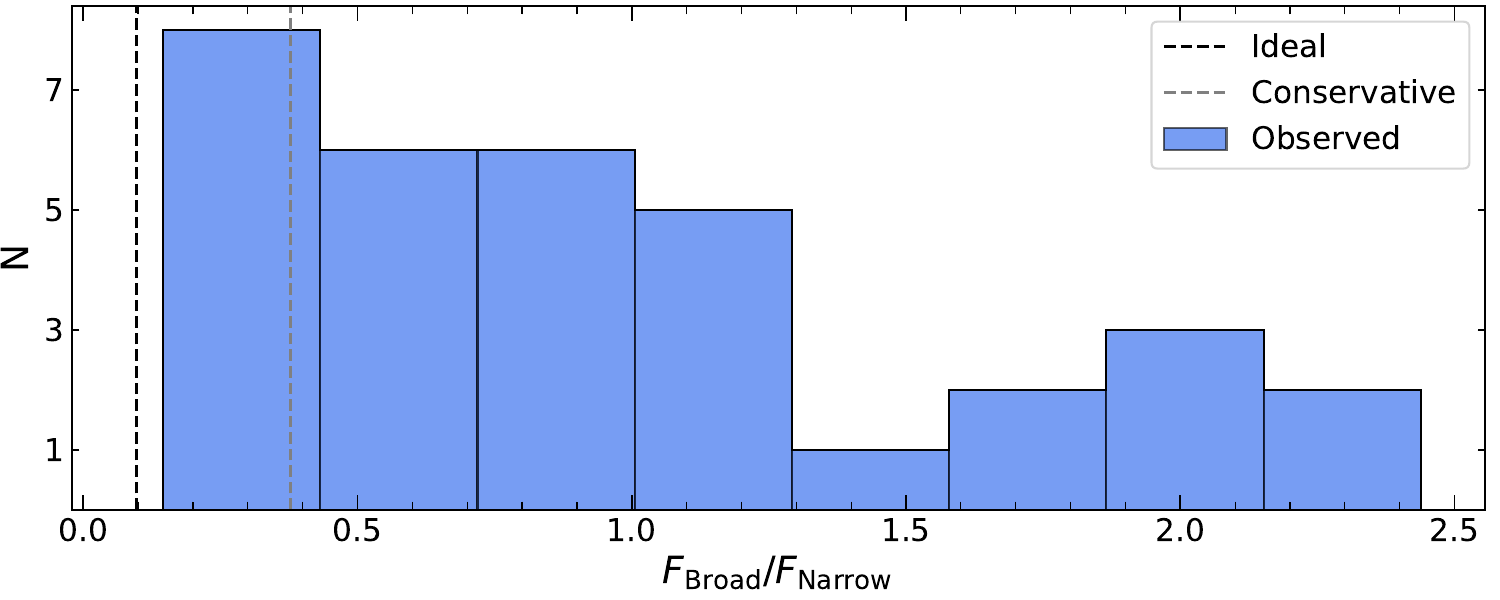}
    \vspace{-0.35cm}
    \caption{The distribution of broad-to-narrow \HA\ flux ratios for galaxies with a detected broad \HA\ emission component. The grey (black) vertical dashed line is the minimum flux ratio required for a broad component to be detected using the fitting procedure in Section \ref{sec:fitting} for a broad component with $\Delta v_{\rm{br}}$ = 10 km s$^{-1}$ ($\Delta v_{\rm{br}}$ = 90 km s$^{-1}$) and a \HA\ signal-to-noise ratio of 25 (100).}
    \label{fig:mock}
\end{figure}

The detection of an underlying broad component depends on several factors. A broad component may be indistinguishable from that of HII regions at low velocities and small velocity offsets. At the same time, low S/N could prevent the detection of a broad component in the faint, high-velocity wings of an emission line, where the broad component may be the strongest. To quantify the detectability of a broad component, we create simulated \HA\ emission lines where we can control the parameters of the broad component and signal-to-noise ratio (SNR) of the line.

For this test, we considered two cases with fixed values for $\Delta v_{\rm{br}}$ and line SNR. In the first case (“conservative”), the broad component had a small velocity offset, $\Delta v_{\rm{br}}$ = 10 km s$^{-1}$, and line SNR = 25, the average value of individual detected galaxies. In the second case (“ideal”), the broad component had a large velocity offset, $\Delta v_{\rm{br}}$ = 90 km s$^{-1}$ and line SNR = 100. For each case, 110 single \HA\ emission lines were simulated, with 11 different FWHM$_{\rm{br}}$ between 300 and 500 km s$^{-1}$ and 10 normalised $A_{\rm{br}}$ between 0.05 and 0.5, which span the range of measured broad components from individual detected galaxies. The values for the simulated narrow components are randomly selected between the measured narrow component values of individual detected galaxies. We adopt the same resolution and wavelength as a \HA\ line of a $z \sim$ 2.3 galaxy. Following the same fitting process and selection criteria as for individual galaxies in Section 2.4.1, we fit the simulated lines with two Gaussians and a single Gaussian and determined which simulated spectra have a detected broad component.

The results of these tests are shown in Figure \ref{fig:mock}. We find that the lowest detectable broad component has an integrated flux of 10\% and 38\% of the narrow component for the “ideal” and “conservative” cases, respectively. Not surprisingly, for individual detections, their broad component flux lies above these limits of the fitting process. For all of the galaxies, their broad flux is greater than the “ideal” limit, while 75\% are above the “conservative” limit. We note, however, that the detection of the broad component is a function of both $\Delta v_{\rm{br}}$ and line SNR. Here, we explored two extreme cases, but varying the values may lead to different limits.

\subsubsection{Composite Spectra}
\label{sec:composite}

Decomposing emission lines into narrow and broad components is often limited by the S/N of a galaxy's spectra. The high-velocity wings of an emission line, where the broad component may be the strongest, are typically dominated by noise. In order to investigate ionised outflows across a wide range of properties, we construct high S/N composite spectra in bins of several galactic properties. We follow a similar method as \cite{Weldon22} for constructing composite spectra. In brief, galaxies in the MOSDEF-ionised sample were grouped together into equal-number bins based on various physical properties (e.g., stellar mass, SFR, \SSFR). The science and error spectra of individual galaxies were shifted to the rest frame, converted to luminosity density, interpolated onto a grid with a wavelength spacing of $\Delta\lambda$ = 0.5\AA, and normalised by either the \OIII\ or \HA\ luminosity, depending on the line of interest, measured from the science spectrum. The composite spectrum at each wavelength point was computed as the weighted average with 3$\sigma$ outlier rejection of the luminosity densities of individual spectra at the same wavelength point, where the weights are 1/$\sigma \left(\lambda \right)^{2}$ and $\sigma \left(\lambda \right)$ is the value of the error spectrum at wavelength $\lambda$. The line-fitting process for the composite spectra is the same as for individual galaxies. The spectra are first normalised using an initial fit, then refitted, the results of which serve as the initial values for an MCMC fitting process. However, when fitting the \HA\ and \NII\ doublet, we include the faint [\ion{S}{II}]$\lambda\lambda$6716,6731 doublet (see Section \ref{sec:loading}). 

\section{Results}
\label{sec:results}

With the decomposition of rest-optical emission lines into narrow and broad components for individual galaxies and stacked spectra in hand, we are in a position to investigate the properties of the ionised outflows against galactic physical properties. The occurrence of ionised outflows is discussed in Section \ref{sec:occurrence}. Section \ref{sec:velocity} focuses on trends between ionised outflow velocity and galactic properties. Section \ref{sec:loading} discusses the mass-loading factor of the ionised outflows and its relations with galactic properties. The energetics of the ionised outflows are discussed in Section \ref{sec:EMloading}.

\subsection{Occurrence}
\label{sec:occurrence}

\begin{figure}
    \includegraphics[width=\columnwidth, keepaspectratio]{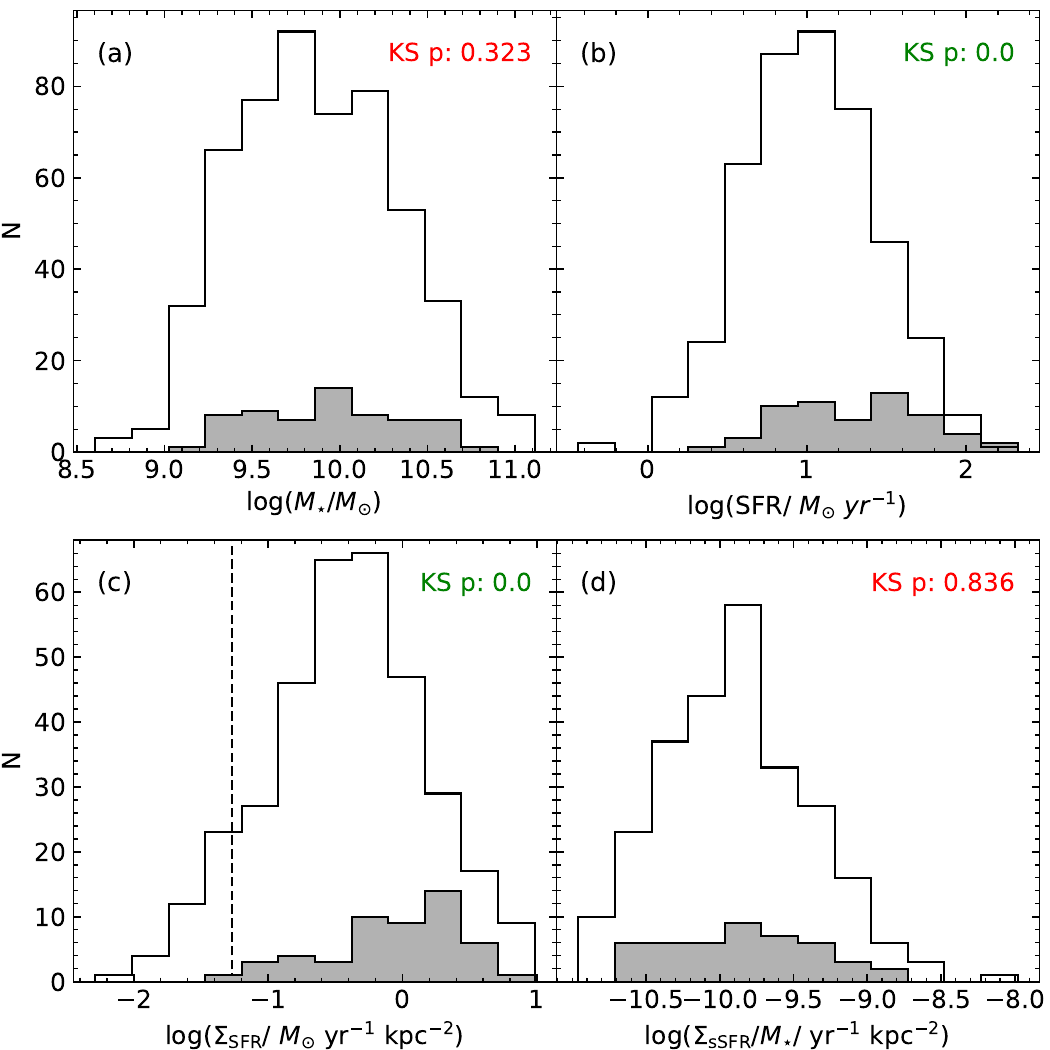}
    \vspace{-0.35cm}
    \caption{The distribution of various galactic properties.  Panel (a): stellar mass, Panel (b): SFR, Panel (c): \SSFR, Panel (d): \sSSFR. Solid grey and open bars represent the 62 galaxies with detected broad components and the remaining galaxies, respectively. The p-value of a KS test between galaxies with a broad component and the remaining galaxies is shown in the upper corners of each panel. The vertical dashed line in panel (c) marks the \SSFR\ threshold proposed by \protect\cite{Heckman02} for launching an outflow.}
    \label{fig:distribution}
\end{figure}

As discussed in Section \ref{sec:selection}, the MOSDEF-ionised sample consists of 598 galaxies, with 392 \OIII\ and 435 \HA\ detections. Of these, there is significant evidence for broad emission in 39 of the 391 (10\%) \OIII\ detections and in 33 of the 435 (7\%) \HA\ detections. Overall, a broad component is detected in 62 (10\%) of the MOSDEF-ionised galaxies. These galaxies typically have a S/N > 50\footnote{We take the S/N as the flux of a single Gaussian divided by the error in the flux for either \OIII\ or \HA.}, highlighting the difficulty of decomposing rest-optical emission lines into multiple components.

\begin{figure*}
  \includegraphics[width=\linewidth, keepaspectratio]{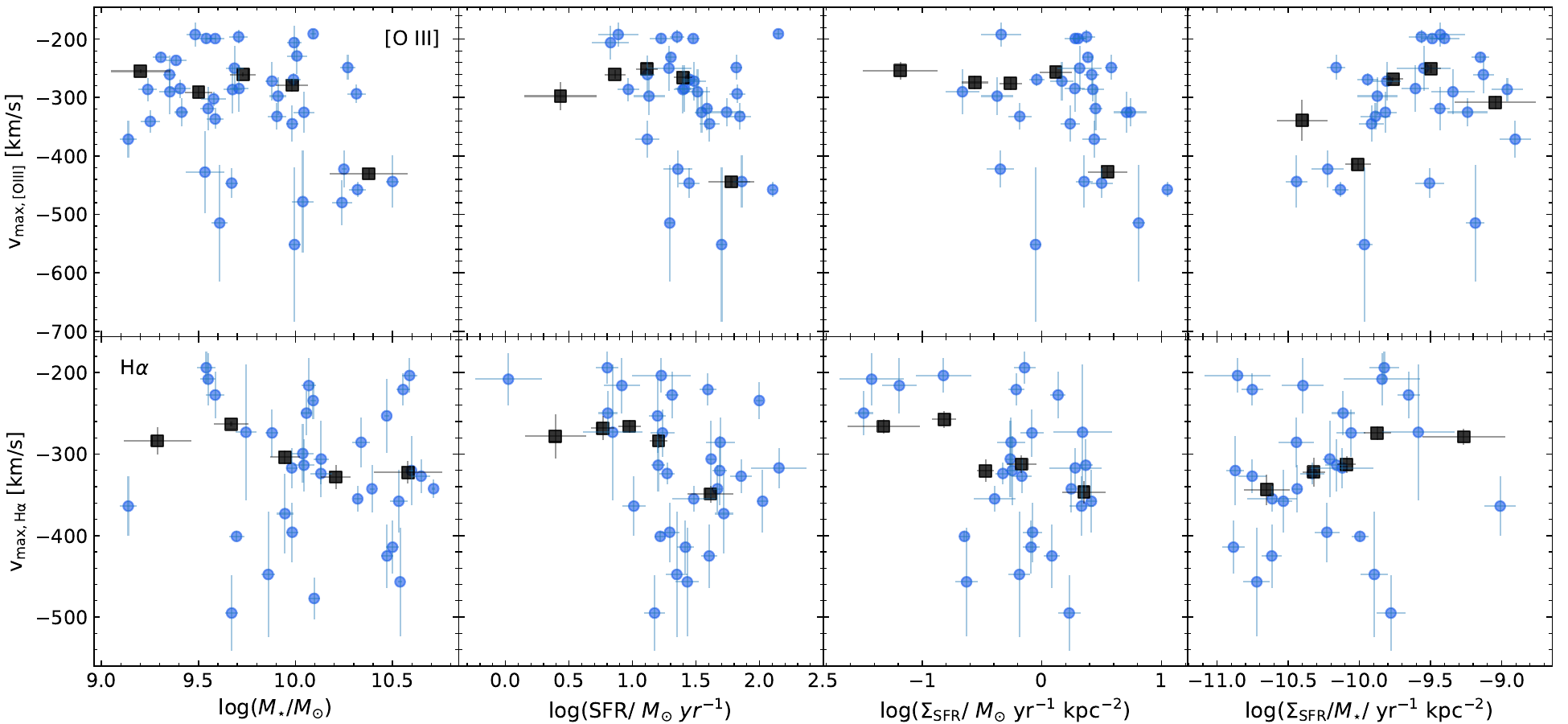}
  \vspace{-0.45cm}
  \caption{Maximum outflow velocity versus various galactic properties. \textit{left}: stellar mass, \textit{left centre}: SFR, \textit{right centre}: \SSFR, \textit{right}: \sSSFR. Top (bottom) rows are plotted versus [\ion{O}{III}] (\HA) \Vmax. Individual galaxies are shown as blue circles, while results from composite spectra are shown as black squares.
  }
  \label{fig:velocity}
\end{figure*}

In Figure \ref{fig:distribution}, we show the distributions of various galactic properties for the subset of galaxies with evidence of ionised outflows and the remaining MOSDEF-ionised galaxies. To quantitatively test whether the galaxies with detected ionised outflows are drawn from the same parent distribution as the remaining galaxies, we perform a Kolmogorov–Smirnov (KS) test. Unsurprisingly, the detection of ionised outflows is strongly tied to signatures of elevated star-formation activity (SFR and \SSFR). This connection is expected as, in star-forming galaxies, the injection of energy and momentum associated with the late stages of massive stellar evolution are theorised to drive galactic-scale outflows. Additionally, nearly all of the galaxies with ionised outflows have an \SSFR\ above the ${\sim} 0.05\ \rm{M_{\odot}}\ \rm{yr}^{-1}\ \rm{kpc}^{-2}$ threshold (dashed-line in panel (c) of Figure \ref{fig:distribution}) proposed by \cite{Heckman02}\footnote{Heckman (2002) propose a 0.1 $\rm{M_{\odot}}\ \rm{yr}^{-1}\ \rm{kpc}^{-2}$ threshold based on local starbursts galaxies. For a Chabrier (2003) IMF assumed in our study, this threshold becomes 0.05 $\rm{M_{\odot}}\ \rm{yr}^{-1}\ \rm{kpc}^{-2}$.}, which is interpreted as the point where energy and momentum can overcome the gravity of the galaxy disk and launch an outflow. Alternatively, the apparent difference between the galactic properties of the outflowing and remaining galaxies may reflect limitations in our fitting technique. Outflows from galaxies with lower SFR or \SSFR\ could be missed if their velocity is low (FWHM$_{\rm{br}}$ < 300 km s$^{-1}$), such that the emission from the broad component is indistinguishable from that of HII regions. Likewise, the detection of a distinct broad emission component likely requires a high SNR of the desired line, and galaxies with strong rest-optical emission lines are also associated with higher star-formation properties. We explore possible dependencies of ionised gas outflow properties on galactic properties in the following sections.

\subsection{Outflow Velocity}
\label{sec:velocity}

A fundamental property of outflowing gas is its velocity. As stellar feedback in star-forming galaxies likely drives their outflows, correlations between outflow velocity and star-formation properties should naturally arise. Several studies have found that the velocity of cool, neutral gas outflows traced by low-ionisation UV absorption lines increases with the star-formation rate and star-formation-rate surface density of the host galaxy \citep[e.g.,][]{Chen10, Steidel10, Martin12, Rubin14, Chisholm15, Heckman15, Bordoloi16, Weldon22}. However, fewer studies have explored how the velocity of warm ionised gas outflows varies with SFR, \SSFR, and other properties of the host galaxy \citep{Swinbank19, Davies19, Avery21, Couto21, Davis23}.

We estimate the maximum outflow velocity from the broad component following previous studies as \Vmax\ = $\Delta v_{\rm{br}}$ $-$ 2$\sigma_{\rm{br}}$, where $\sigma_{\rm{br}}$ is the Gaussian sigma value of the broad component \citep[see][]{Genzel11, Genzel14, Wood15}. Figure \ref{fig:velocity} presents \Vmax\ derived from \OIII\ and \HA\ as a function of stellar mass, SFR, \SSFR, and \sSSFR. Considering individual galaxies (blue circles), we find no significant correlations between \Vmax\ and properties of the host galaxy, which may be due to limitations in our fitting technique (Section \ref{sec:dyn_range}). On the other hand, the composite spectra (black squares) exhibit clear trends with velocity. The highest stellar mass, SFR, and \SSFR\ bins appear to have gas at significantly larger  $V_{\rm{max, [\ion{O}{III}]}}$ compared to the lowest bins. Similarly, $V_{\rm{max, H\alpha}}$ appears faster in the higher (lower) \SSFR\ (\sSSFR) bin compared to the lowest (highest) bin. The sudden increase in $V_{\rm{max, [\ion{O}{III}]}}$ in the highest galactic properties bins may reflect different initial conditions of the outflowing gas. The \OIII–emitting zone is likely more compacted compared to the HII region, due to the higher ionisation potential of [\ion{O}{II}] than HI – 35 eV and 13.6 eV, respectively. Removing the \OIII\ gas from a smaller, more tightly bound region then likely requires more extreme conditions (i.e., higher SFR, \SSFR). The trend of galaxies with higher SFR and \SSFR\ hosting faster outflows agrees with the picture of outflows driven by feedback from star-formation. Although, these global trends are quite weak, with $V_{\rm{max, H\alpha}} \propto \Sigma_{\rm{SFR}}^{0.07\pm0.03}$. On smaller 1 -- 2 kpc scales, \cite{Davies19} measured broad \HA\ components in composites of IFU \HA\ observations from 28 $z$ $\sim$ 2.3 galaxies and found \SSFR\ and outflows are closely, with $v_{\rm{out}} \propto \Sigma_{\rm{SFR}}^{0.34\pm0.10}$. Conversely, the $V_{\rm{max, H\alpha}}$ trend with \sSSFR\ suggests that – at a fixed \SSFR\ – faster outflows are launched from high-mass galaxies relative to low-mass ones. However, the observed trend is quite weak, with $V_{\rm{max, H\alpha}}$ increasing only by $\sim$80 km s$^{-1}$ across the roughly 1.5 dex range in \sSSFR\ for the composite spectra.

\subsection{Mass-Loading Factor}
\label{sec:loading}

In this section, we turn towards estimating the mass-loading factor of the ionised outflows, focusing on galaxies with a detected broad \HA\ component. While a similar analysis is possible for the \OIII\ line, there are additional dependencies (i.e., the chemical enrichment of the outflowing gas) that further complicate the derived values. Previous studies have found that the mass-loading factor derived from \OIII\ is consistent with but systematically lower than values derived from \HA\ \citep[see][]{Carniani15, Marasco20, Concas22}.

Adopting the simple outflow model described in \cite{Genzel11} and \cite{Newman12}, we estimate the mass outflow rate ($\dot{M}_{\rm{out}}$) of the galaxies. This model is based on three main assumptions: (1) the geometry of the outflow is multi-conical or spherical with constant velocity and mass loss, (2) the gas in the broad component is photoionised and in case B recombination with an electron temperature of $T_{e}$=10$^{4}$K, and (3) the electron density of the broad component does not vary significantly with radius. Under these conditions the mass outflow rate can be calculated as: 
\begin{equation}
    \label{equ:mdot}
    \dot{M}_{\rm{out}} = \frac{1.36m_{H}}{\gamma_{\text{H}\alpha} n_{e}} \left( L_{\text{H}\alpha,\  \rm{Broad}}\right) \frac{V_{\rm{out}}}{R_{\rm{out}}}
\end{equation}
where $m_{H}$ is the atomic mass of hydrogen, $\gamma_{\text{H}\alpha}$($T_{e}$) = 3.56$\times$10$^{-25}T_{4}^{-0.91}$ erg cm$^{-3}$ s$^{-1}$ is the \HA\ emissivity at an electron temperature $T_{4}$ = 10$^{4}$K, $n_{e}$ is the electron density of the outflow, $L_{\text{H}\alpha, \rm{Broad}}$ is the extinction corrected \HA\ luminosity of the broad component, $V_{\rm{out}}$ is the velocity of the outflow, and $R_{\rm{out}}$ is the radial extent of the outflow.

We adopt this model as the observations are not resolved and to facilitate comparisons with similar studies in the literature. However, varying the assumptions of the model can have noticeable effects on \Mdot. For example, the inverse dependence of $\gamma_{\text{H}\alpha}$ on electron temperature creates a temperature dependence for \Mdot. \cite{Sanders20} preformed direct measurements of $T_{e}$ from the [\ion{O}{III}]$\lambda$4363 line for four star-forming galaxies in the  MOSDEF survey. They found an ISM $T_{e}$ of $\sim$14000 $-$ 17000 K, with a value of 15400 K for a composite of the four galaxies. If we assume this higher electron temperature for the outflowing gas, \Mdot\ would increase by a factor of 1.5. At higher temperatures, the gas responsible for the broad component may be dominated by collisional ionisation, rather than photonised. \cite{Genzel11} considered collisional ionisation at $T_{e}$ = 2$\times$10$^{4}$ K and found that the inferred \Mdot\ decreased by a factor of 2, with the difference decreasing at lower temperatures.

In principle, the electron density of the outflowing gas can be calculated from the broad [\ion{S}{II}]$\lambda$6717/[\ion{S}{II}]$\lambda$6732 line ratio \citep{Osterbrock89, Sanders16}. Using the composite spectra, we attempt to measure this ratio directly. However, the broad components of the [\ion{S}{II}] doublet are not well constrained in any of the composite spectra. Instead, we follow a similar approach as \cite{Schreiber19} and search for broad components in a composite spectrum of the 33 galaxies with detected broad \HA\ components. In this new composite, the broad components of the [\ion{S}{II}] doublet are well constrained. Using the relationship from \cite{Sanders16}, we estimate an electron density of 420$_{-200}^{+260}$ cm$^{-3}$, consistent within the large uncertainties of the 300 -- 500 cm$^{-3}$ range of $n_{e, \text{broad}}$ reported for ionised outflows in local and high-$z$ star-forming galaxies \citep[e.g.,][]{Arribas14, Ho2014, Schreiber19, Fluetsch21}. The intrinsic \HA\ luminosity of the broad component is measured individually for each galaxy by scaling the total corrected \HA\ luminosity by the broad-to-single \HA\ ratio (F$_{\text{broad}}$/F$_{\text{Single}}$). For $V_{\text{out}}$, we adopted the maximum velocity from Section \ref{sec:velocity}. We take the radial extent of the outflows to be their hosts' effective radii, $R_{\rm{out}}$ = \RE, motivated by high-resolution adaptive optics SINFONI observations of ionised gas outflows in high redshift star-forming galaxies indicating that outflows typically extend over the half-light radius \citep{Newman12, Schreiber14}.

\begin{figure*}
  \includegraphics[width=\linewidth, keepaspectratio]{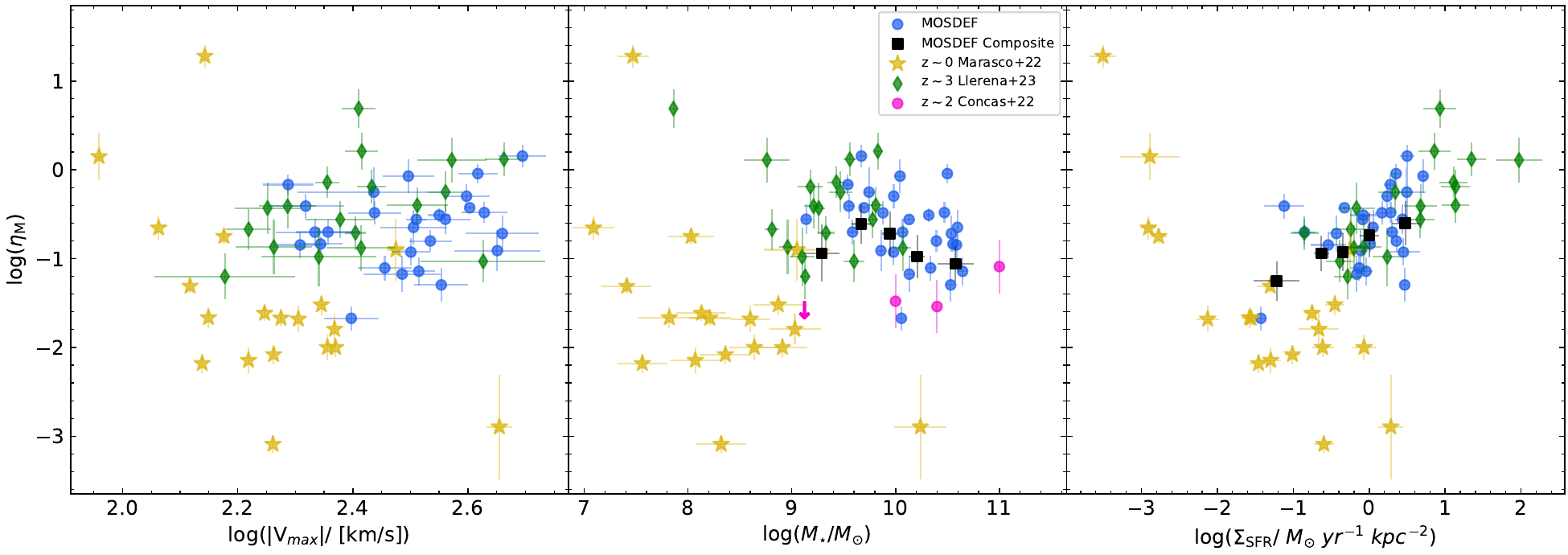}
  \vspace{-0.45cm}
  \caption{Mass-loading factor versus various galactic properties, compared with values from the literature. \textit{left}: log(|V$_{\rm{max}}$|), \textit{centre}: log(\Mste), \textit{right}: log($\Sigma_{\rm{SFR}}$). Individual MOSDEF galaxies are shown as blue circles, while results from composite spectra are shown as black squares. Markers show observational results including low-mass star-forming galaxies at $z \sim$ 3 from \protect\citet[][green diamounds]{Llerena23}, local starburst dwarf galaxies from \protect\citet[][yellow stars]{Marasco23}, and composites of  $z \sim$ 2 star-forming galaxies from \protect\citet[][pink circles]{Concas22}.
  }
  \label{fig:mass loading}
\end{figure*}

The mass-loading factor (\Mload) represents the amount of mass removed by an outflow per stellar mass formed, and, for star-formation driven outflows, is thought of as a diagnostic of outflow efficiency. Specifically, \Mload\ is defined as the outflow rate normalised by the star formation rate: \Mload\ = $\dot{M}_{\rm{out}}$/SFR. An estimate of the SFR can be obtained from the narrow Gaussian component of \HA\ tracing the ongoing star formation activity in the galaxy disk: 
\begin{equation}
    \text{SFR}_{\text{Narrow}} = 3.236\times10^{-42} \left( L_{\text{H}\alpha} \frac{F_{\text{Narrow}}}{F_{\text{Single}}}\right) \left[M_{\odot}/yr \right]
\end{equation}
where the first term is the conversion factor between \HA\ luminosity and SFR from \cite{Reddy18b} (Section \ref{sec:gal_props}) and $F_{\text{Narrow}}$/$F_{\text{Single}}$ is the fraction of the total flux in the narrow component. When divided by the narrow component \HA\ SFR and simplified, Equation \ref{equ:mdot} can be written as 
\begin{equation}
    \label{equ:mload}
    \eta_{m} \approx 3.05\left(\frac{100\ \text{cm}^{-3}}{n_{e}}\right) \left(\frac{V_{\text{max}}}{300\ \text{km}\ \text{s}^{-1}}\right) \left(\frac{\text{kpc}}{R_{\text{E}}}\right) \left(\frac{F_{\text{Broad}}}{F_{\text{Narrow}}}\right)
\end{equation}
where $F_{\text{Broad}}$/$F_{\text{Narrow}}$ is the broad-to-narrow flux ratio. Table \ref{tbl:HA} lists the mass-loading factor for galaxies with a detected broad \HA\ component, finding \Mload\ ranging from 0.02 -- 1.44, with a median of 0.23. Uncertainties on \Mload\ are taken as the dispersion of 1000 realisations after perturbing \Vmax, \RE, and $F_{\text{Broad}}$/$F_{\text{Narrow}}$ by their errors. We do not include errors in the electron density and temperature assumed, and including these errors would increase the error on \Mload\ by $\sim$0.2 dex. 

The values we estimate for \Mload are similar to those of other recent studies of ionised outflows. \cite{Swinbank19} stacked \HA\ emission of $\sim$530 star-forming galaxies at $z \sim$ 1 and found \Mload\ $\sim$ 0.1--0.4. Similarly, at $z \sim$ 2, \cite{Davies19} and \cite{Schreiber19} stacked \HA\ emission from star-forming galaxies finding \Mload\ $\sim$  0.3--0.5 and \Mload\ $\sim$ 0.1 -- 0.2, respectively. Recently, \cite{Llerena23} investigate ionised outflows in a sample of low-mass (7.8 < log($M_{\star}/M_{\odot}$) < 10.2) star-forming galaxies at $z \sim$ 3 and found a wide range of \Mload, from 0.1 to 4.9. However, there is debate about the typical value of \Mload\ in star-forming galaxies, with some studies reporting values less than 0.1 \citep{Concas22, Marasco23}. The tension in \Mload\ between various studies is likely due to different assumptions when estimating $\dot{M}_{\rm{out}}$, or differences in the methodology used to measure ionised outflows \citep[see discussion in][]{Concas22}.

\subsubsection{Trends}
\label{sec:loading_trend}

In this section, we explore which, if any, internal galaxy properties correlate with the mass-loading factor. Of particular interest is how \Mload\ potentially scales with the (1) ionised outflow velocity, (2) stellar mass, and/or (3) \SSFR. Regarding the first two properties, simple analytical arguments and numerical simulations predict an anti-correlation between the mass-loading factor and the outflow velocity or stellar mass of galaxies, suggesting that outflows are more efficient at removing material from the shallower potential wells of lower-mass galaxies \citep{Murray05, Oppenheimer08, Muratov15}. Additionally, the scaling relation between \Mload\ and outflow velocity or stellar mass is predicted to depend on the driving mechanism of the outflow, which scales steeper if the outflows are energy-driven and shallower if they are momentum-driven (see Section \ref{sec:dirving}). At the same time, one might expect a correlation with \SSFR, which traces the concentration of star formation in a galaxy, as regions with higher \SSFR\ will be more efficient at injecting energy and momentum into the ISM from overlapping supernovae or stellar winds from massive stars, resulting in conditions amenable for outflows.

Figure \ref{fig:mass loading} presents the variation of the mass-loading factor as a function of \Vmax, stellar mass, and \SSFR\ with similar results from the literature. In the MOSDEF-ionised sample, we find that \Mload\ is not significantly correlated with \Vmax, while marginally correlated with both stellar mass (2.5$\sigma$) and \SSFR\ (2.2$\sigma$). Galaxies with lower masses appear to have larger \Mload\ than higher mass galaxies, with \Mload\ decreasing by a factor of 0.6 over a mass range log($M_{\star}/M_{\odot}$) = 9 -- 10.7. Although, at a fixed stellar mass, there is a large range in \Mload, which may reflect the variation in outflow efficiency amongst different outflow phases for individual galaxies. On the other hand, galaxies with \textit{higher} \SSFR\ appear to have larger \Mload\ than lower \SSFR\ galaxies, with \Mload\ increasing 1.8 dex over an \SSFR\ range log(\SSFR/$M_{\odot}$ yr$^{-1}$ kpc$^{-2}$) = -1.4 -- 0.7.

A negative relation between \Mload\ and stellar mass is predicted by analytical arguments and simulations. Observations, however, have yielded ambiguous results, as shown in Fig \ref{fig:mass loading}. Studies that measure ionised outflows in composite spectra tend to find a roughly constant relation between \Mload\ and stellar mass, while studies of individual galaxies generally find the expected negative correlation between \Mload\ and stellar mass. At $z \sim$ 2, \cite{Concas22} reported relatively low mass-loading factors that exhibit little variation among three stellar mass bins below 10$^{11} M_{\odot}$, but increases $\sim$0.2 dex towards the highest mass bin – likely due to increased AGN activity in more massive systems. Meanwhile, \cite{Llerena23} and \cite{Marasco23} found that \Mload\ decreases with stellar mass for individual $z \sim$ 3 star-forming and local dwarf starburst galaxies, respectively. However, the mass-loading factors of \cite{Marasco23} are offset towards significantly lower values compared to the MOSDEF-ionised sample. This discrepancy may be due to differing methodologies used to parameterise the outflow velocity, varying assumptions about the outflow geometry, or it may suggest a possible redshift evolution of the mass-loading factor.

The correlation between \Mload\ and \SSFR\ agrees with the simple outflow picture: galaxies with higher \SSFR\ are more effective in driving outflows. However, this relation is likely caused by an underlying correlation between \SSFR, \Mload, and \RE. With our adopted definitions, \SSFR\ $\propto$ \RE$^{-2}$ and \Mload\ $\propto$ \RE$^{-1}$, thus a positive relation would naturally arise. A similar positive relationship is found by \cite{Llerena23} who also derived \SSFR\ and \Mload\ using effective radii. Conversely, \cite{Marasco23} derived \Mload\ with outflow radii measured directly from their sample – independent of the effective radius – and found a strong negative correlation between \Mload\ and \SSFR.

\subsection{Outflow Energetics}
\label{sec:EMloading}

\begin{figure}
    \includegraphics[width=\columnwidth, keepaspectratio]{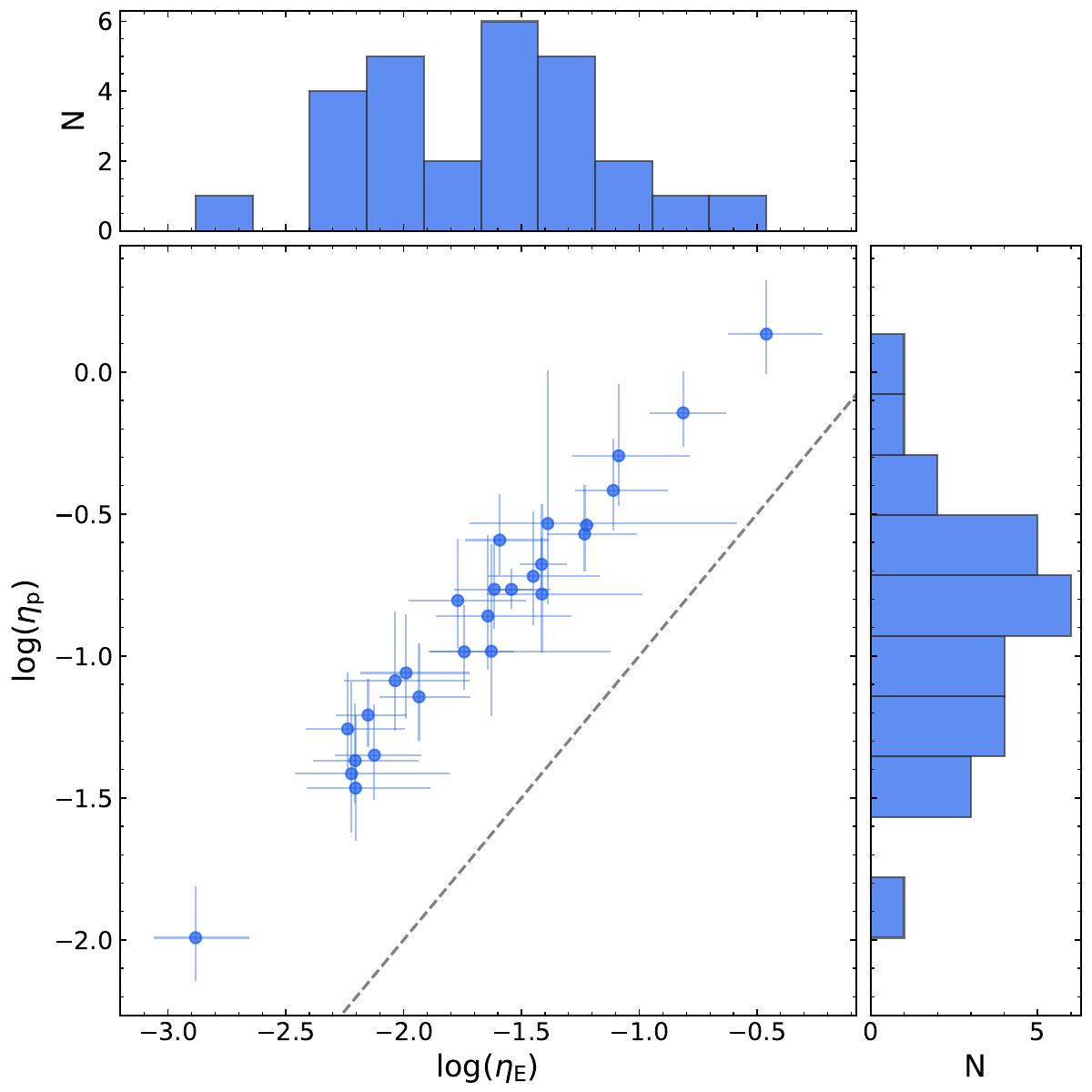}
    \vspace{-0.45cm}
    \caption{Momentum-loading factor versus energy-loading factor. The dashed line marks the one-to-one line. In all of the galaxies, the outflows appear to carry away more momentum than kinetic energy from the ISM.}
    \label{fig:energy-momentum}
\end{figure}

In addition to mass, outflows are characterised by the amount of energy and momentum they remove from the ISM. In Appendix \ref{sec:derivation}, we present a derivation for the energy and momentum outflow rates and reference rates of energy and momentum produced within galaxies. Briefly, the energy and momentum outflow rates are calculated from the mass outflow rate (Equation \ref{equ:mdot}). Reference rates are calculated following similar analytic arguments as \cite{Murray05}, focusing on the injection of energy and momentum from type IIa supernovae. Dividing the outflow rates by the reference rates and simplifying, the energy- and momentum-loading factors of the outflowing ionised gas are given by:
\begin{equation}
      \eta_{E} = 0.27\left(\frac{100\ \text{cm}^{-3}}{n_{e}}\right) \left(\frac{V_{\text{max}}}{300\ \text{km}\ \text{s}^{-1}}\right)^{3} \left(\frac{\text{kpc}}{R_{\text{E}}}\right) \left(\frac{F_{\text{Broad}}}{F_{\text{Narrow}}}\right)
      \label{equ:etaE}
\end{equation}
\begin{equation}
      \eta_{p} = 1.76\left(\frac{100\ \text{cm}^{-3}}{n_{e}}\right) \left(\frac{V_{\text{max}}}{300\ \text{km}\ \text{s}^{-1}}\right)^{2} \left(\frac{\text{kpc}}{R_{\text{E}}}\right) \left(\frac{F_{\text{Broad}}}{F_{\text{Narrow}}}\right)
      \label{equ:etaP}
\end{equation}
Directly comparing Equation \ref{equ:etaE} and \ref{equ:etaP}, we find that the ratio of the momentum- to energy-loading factor is simply:
\begin{equation}
    \frac{\eta_{p}}{\eta_{E}} = 6.5\left( \frac{300\ \text{km}\ \text{s}^{-1}}{V_{\text{max}}} \right)
\end{equation}

For the sample, \Vmax\ ranges between $\sim$200 -- 500 km s$^{-1}$, suggesting that these ionised outflows carry away more momentum than kinetic energy from the ISM, with \Pload/\Eload\ ranging from $\sim$4 to 10. We further test this result by varying the various physical parameters within a factor of 2 from their adopted values (see Appendix \ref{sec:derivation}). Considering the most favourable case, \Pload\ remains larger than \Eload\ by about a factor of 2. 

Taken at face value, the larger momentum-loading factors suggest that the ionised outflows are primarily momentum driven. However, the comparison between momentum- and energy-loading factors is highly model dependent. There are other sources of energy and/or momentum within the galaxy not accounted in our analytic reference rates (e.g., cosmic rays, winds from massive stars), such that \Eload\ and \Pload\ are likely upper limits. For example, in Figure \ref{fig:energy-momentum}, there are a few galaxies with abnormally high energy- and momentum-loading factors (\Eload\ > 0.1, \Pload\ $\sim$ 1), which would suggest that their ionised outflows remove nearly all of the energy and momentum produced.\footnote{High-resolution hydrodynamical simulations find that the energy from supernovae is rapidly thermalised and radiated away, with effectively $\sim$10\% transferred to the ISM to drive outflows \citep{Creasey13}.} Additionally, the assumptions made in our calculations may be incorrect. Based on high-resolution “local patch” simulations of supernovae in the ISM, \cite{Kim15} argue that the momentum injected by supernovae is better represented by the spherical momentum at the end of the Sedov-Taylor stage when an SN blast wave cools and a shell forms, rather than the initial injected momentum. If we adopted this convention, \Pload\ decreases by $\sim$0.72 dex, such that \Pload\ and \Eload\ are about equal. In Section \ref{sec:dirving}, we further investigate the primary driving mechanism of the ionised outflows.

\section{Discussion}
\label{sec:discussion}

\subsection{Driving Mechanism}
\label{sec:dirving}

While there is a general picture of the origin of outflows and their role in galaxy evolution, the physical mechanism(s) that generates and sustains outflows remains an open question. In star-forming galaxies, outflows could be launched by momentum injected into the ISM by supernovae, or by radiation pressure acting on dust grains accelerating gas coupled to the dust \citep["momentum$-$driven";][]{Murray05, Murray11}. In addition to injecting momentum, mechanical energy released from multiple, overlapping supernovae thermalises a large fraction of nearby gas forming a hot over-pressured bubble that sweeps up ambient ISM material until it is ejected from the galaxy \citep["energy$-$driven";][]{Chevalier85}. Finally, cosmic rays produced by supernovae can transfer momentum to gas after scattering off of magnetic inhomogeneities in the ISM as they diffuse out of a galaxy \citep[see discussions in][]{Heckman17, Zhang18}. Simple analytic arguments predict that for purely momentum-driven outflows the mass-loading factor scales as \Mload\ $\propto V_{\text{out}}^{-1}$ and \Mload\ $\propto M_{\star}^{-1/3}$. Similarly, for purely energy-driven outflows: \Mload\ $\propto V_{\text{out}}^{-2}$ and \Mload\ $\propto M_{\star}^{-2/3}$ \citep{Murray05}. As these mechanisms are likely dominate under different galactic conditions, outflows could be driven by a combination of mechanical energy, radiation pressure, and cosmic rays. 

\begin{figure}
    \includegraphics[width=\columnwidth, keepaspectratio]{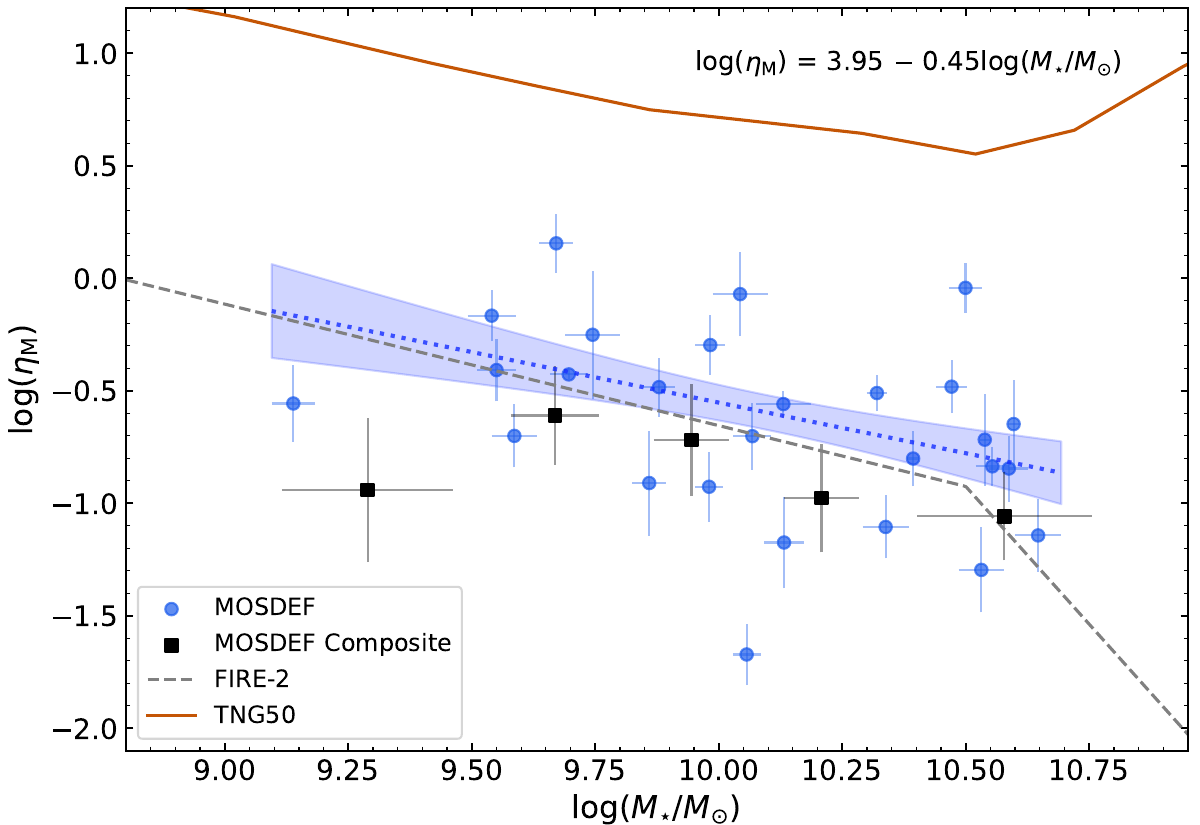}
    \vspace{-0.45cm}
    \caption{Mass-loading factor as a function of stellar mass, compared with predictions from simulations. Individual MOSDEF galaxies are shown as blue circles, while results from composite spectra are shown as black squares. 
    The dotted blue line and shaded region (68\% confidence intervals) is the best-fit line to the MOSDEF-ionised galaxies. The functional form of the line is listed in the upper-right corner. Lines show theoretical predictions from the FIRE-2 \protect\citet[][dashed grey]{Pandya21} and Illustris TNG50 \protect\citet[][solid orange]{Nelson19} cosmological simulations.}
    \label{fig:eta-fit}
\end{figure}

Here, we investigate the marginally-correlated trend of \Mload with stellar mass. Figure \ref{fig:eta-fit} presents the best-fit linear regression in logarithmic space for galaxies with detected \HA\ outflows:
\begin{equation}
    \text{log}(\eta_{\text{M}}) = (3.95 \pm 0.54) - (0.45 \pm 0.06)\text{log}(M_{\star}/M_{\odot}) 
\end{equation}
The power-law index of -0.45 is intermediate between the $M_{\star}^{-2/3}$ and $M_{\star}^{-1/3}$ dependence predicted for energy- or momentum-driven outflows, suggesting that these ionised outflows are driven by a combination of mechanical energy and radiation pressure. 

In Figure \ref{fig:eta-fit}, we also show the comparison between our derived \Mload--\Mste\ relation to theoretical predictions from Illustris TNG50 \citep{Nelson19, Pillepich19} and Feedback in Realistic Environments \citep[FIRE;][]{Hopkins14, Hopkins18, Hopkins23} cosmological simulations. In particular, we focus on the TNG50 values derived from outflowing gas at a fixed distance of 10 kpc from the galaxy with a radial velocity > 0 km s$^{-1}$ and the FIRE-2 values for warm (10$^{3}$ < T < 10$^{5}$ K) outflowing gas at a fixed thickness of 0.1--0.2$R_{\text{vir}}$\footnote{The average $R_{\text{vir}}$ of our sample is 100 kpc -- estimated using the stellar-to-halo mass relation of \cite{Behroozi19}} \citep{Pandya21}. We note that this is not a direct comparison, as the simulations measure \Mload\ at $\sim$4$\times$ larger distances than the effective radii used in our derivation of \Mload. As the properties of the outflowing gas (e.g., density, velocity) change with distance, \Mload\ would also vary with distance. For example, \cite{Nelson19} measured lower \Mload\ at larger distances (see their Figure 5).

Below log($M_{\star}/M_{\odot}$) $\approx$ 10.5, there is general agreement of a negative correlation between stellar mass and \Mload. Unsurprisingly, the theoretical values from TNG50 are larger than the observed MOSDEF-ionised mass-loading factors. In these types of large-volume simulations, small scales are not resolved, instead relying on sub-grid recipes to describe stellar feedback, which may overpredict the efficiency of stellar feedback. Additionally, the TNG50 values represent the total mass-loading factor of all phases, rather than our values which only trace ionised outflows from rest-optical emission lines. On the other hand, there is remarkable ($\sim$1$\sigma$) agreement between our derived relation and the theoretical prediction from FIRE-2. \cite{Pandya21} reported a broken power law dependence between \Mload\ and stellar mass, with a shallower (-0.54$\pm$0.05) slope below and steeper (-2.45$\pm$0.3) slope above log($M_{\star}/M_{\odot}$)$\sim$10.5. Although, at log($M_{\star}/M_{\odot}$) $\sim$ 10.5, we do not see strong evidence for a sudden drop in \Mload. The broken power law relation may be due to decreasing ISM resolution in the FIRE-2 simulations towards higher stellar masses, such that \Mload\ is underestimated. However, we caution our derived relation is determined from a \textit{marginal} correlation between \Mload\ and stellar mass. 

\subsection{Fate and impact of ionised outflows}
\label{sec:escape}

The impact of outflows on their hosts’ evolution is directly related to the amount of material that escapes from the gravitational potential well of their host. Here, we investigate whether the ionised outflows have sufficient speeds to escape the gravitational potential of their host or whether the gas is retained and likely recycled as a part of a galactic fountain. For an isothermal gravitational potential truncated at $r_{\text{max}}$, the escape velocity at radius $r$ is
\begin{equation}
    v_{\text{esc}}(r) = v_{\text{circ}}\sqrt{2 \left[1 + \text{ln}\left(r_{\text{max}}/r \right) \right]}
    \label{equ:escape}
\end{equation}
where \Vcir\ is the circular velocity of the galaxy, taken from \cite{Price20}. Briefly, for galaxies with resolved and detected rotation measured from their 2D spectra, circular velocities are calculated as $v_{\rm{circ}}(1.3R_{\rm{E}}) = \sqrt{V(1.3R_{\rm{E}})^{2} + 4.4\sigma_{V,0}^{2}}$, where $\sigma_{V,0}$ is the intrinsic galaxy velocity dispersion.\footnote{We take $r$ = 1.3$R_{\rm{E}}$ for the radius for galaxy's circular velocity as a conservative choice, as this is the radius where an exponential rotation curve peaks.} Otherwise, circular velocities are inferred using integrated velocity dispersions and the best-fit ensemble $V$/$\sigma$ from galaxies without detected rotation.

\begin{figure}
    \includegraphics[width=\columnwidth, keepaspectratio]{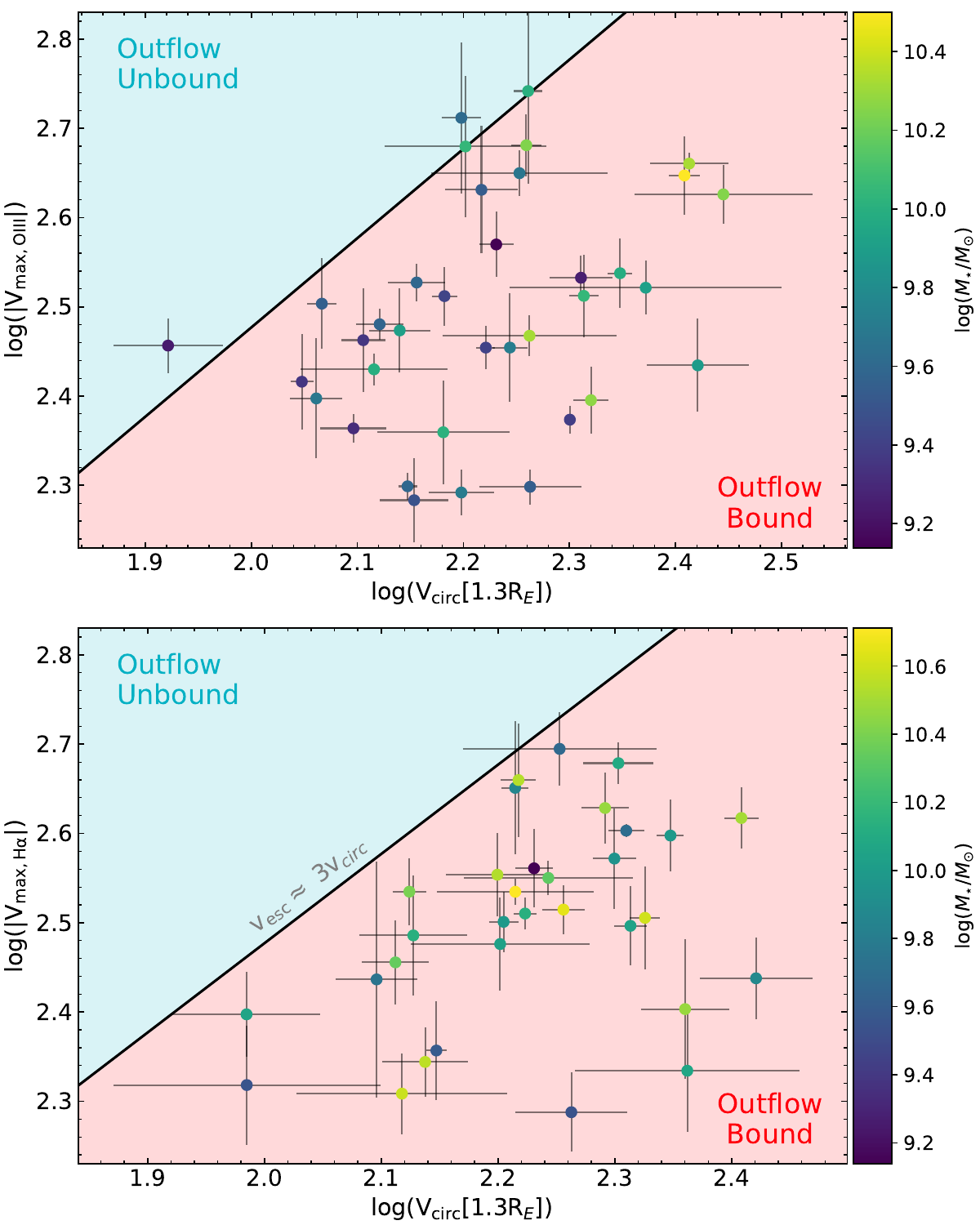}
    \vspace{-0.45cm}
    \caption{Maximum ionised outflow velocity as a function of circular velocity. \textit{Top}: \OIII\ \textit{Bottom}: \HA. The line denotes the gas velocity required to escape the gravitational potential assuming an isothermal gravitation potential that extends to a maximum radius of $r_{\rm{max}}$, see Equation \ref{equ:escape}. Outflowing gas above 3\Vcir\ ($r_{\rm{max}}$ / $r$ = 33) likely has enough velocity to escape, while below 3\Vcir\ the gas is likely retained. Objects are colour coded according to their stellar mass.}
    \label{fig:escape}
\end{figure}

Figure \ref{fig:escape} shows the maximum ionised outflow velocity as a function of circular velocity. 
To gauge whether the gas escapes, we adopt a rather conservative threshold of $r_{\text{max}}$/$r$ = 33 (\Vesc\ = 3\Vcir), such that outflowing gas likely has enough velocity to escape or is retained \citep{Veilleux05, Veilleux20}. In the top panel, we find that outflows detected in \OIII\ appear to escape from four and are retained by 32 galaxies, while outflows detected in \HA\ are retained in all of the galaxies. In addition, we do not find that outflows from shallower potential wells lie closer to the \Vesc\ = 3\Vcir\ line (i.e., more easily removed) compared to outflows from deeper potential wells. It thus seems that the ionised outflows studied in this work are predominantly retained, recycling back onto the galaxy as part of a galactic fountain, rather than escaping into the IGM. However, these outflows could still potentially contribute to heating the CGM, reducing the rate at which gas can accrete and suppressing star formation.

Multi-wavelength observations and multi-phase simulations have investigated the contribution of ionised outflows to the total mass and energy outflow rates. On the observation side, studies on molecular, neutral, and ionised outflows in local AGNs, ULIRGs, and starburst galaxies \citep[e.g.,][]{Cicone14, Carniani15, Leroy15, Rupke17, Fluetsch19, Fluetsch21} have found that the molecular and neutral phases typically dominate the mass outflow rate. Similarly, local patch and zoom-in simulations predict that the majority of outflowing energy (mass) is carried in the hot (cold) phase \citep[][]{Kim18, Kim20, Pandya21}. In the MOSDEF-ionised sample, the high fraction of warm, ionised outflows retained and their modest mass-, energy-, and momentum-loading factors suggest their contribution may be negligible, even during the peak of cosmic star-formation. However, without observations of other outflow phases (i.e., hot, neutral, molecular), we cannot constrain the contribution of these outflows to the total mass, energy, and momentum outflow rates.

\subsection{Significance of Outflow Velocity and Galactic Properties}
\label{sec:turbulent}

As discussed in Section \ref{sec:occurrence}, in individual galaxies, neither the maximum ionised outflow velocity derived from broad \OIII\ or \HA\ emission components appear to correlate significantly with any galactic property. In particular, the lack of a relation between outflow velocity and star formation properties appears to be in tension with the picture of stellar feedback driven outflows, as the level of star formation activity should set the amount of energy and momentum injected into the ISM. However, this apparent lack of observed relations may be due to contributions to the broad component from turbulent motions and/or the limited dynamic range of properties probed by galaxies with detected broad components in the MOSDEF sample.

\subsubsection{Turbulence}

Throughout this paper, we have adopted the interpretation that the broad component of rest-optical emission lines is a tracer of ionised gas entrained in star-formation driven outflows. However, the broad emission component may originate from other sources. Here, we consider whether shocks can explain the observed broad emission components within the sample.

Outflows can produce widespread shocks throughout a galaxy by injecting mechanical energy into the ISM. As shocked regions have high electron temperatures and ionisation states \citep[e.g.,][]{Dopita96}, collisional excitation and ionisation from shocks can produce a variety of optical emission lines, creating broad emission line components. In slow shocks (V < 200 km s$^{-1}$), the shock front moves faster than the photoionisation front, producing relatively weak high ionisation lines, but strong low ionisation lines. Conversely, in fast shocks (V > 200 km s$^{-1}$) a supersonic photoionisation front pre-ionises the gas – known as a precursor – which produces strong high ionisation lines. The emission line ratios of shocked gas differ from those of gas photoionised in HII regions, often with higher \NII/\HA\ and \OIII/\HB\ ratios similar to gas photoionised by AGN \citep[e.g.,][]{Allen08, Alarie19}.

If the broad components arise from shocked gas, the width of the broad component would trace the velocity of the shock rather than the velocity of outflowing material. To investigate whether shocks can explain the broad components, we consider MAPPINGS V \citep{Sutherland17, Sutherland18} fully radiative shock models from the 3MdBs\footnote{\url{http://3mdb.astro.unam.mx/}} database \citep{Alarie19}. This database provides simulated emission line ratios for shock gas, the precursor, and a combination of shock and precursor for multiple grids. To facilitate comparisons between our observations and the models, we use the low metallicity grid, matched to the broad range of metallicities, 6.64 < 12+log(O/H) < 9.28, used by \cite{Gutkin16}. This grid spans a large range in shock parameters -- for each metallicity -- with shock velocities of $V_{s}$ = 100 -- 1000 km s$^{-1}$, magnetic field parameters of B$_{\text{o}}$ = 10$^{-4}$ -- 10 $\mu$G, and pre-shock densities of n$_{\text{o}}$ = 1 -- 10000 cm$^{-3}$.

For this analysis, we focus on the composite spectra binned by stellar mass, as the broad components of \HB\ and \NII\ are not robustly detected in individual galaxies, and mass is estimated independently from the \HA\ line, thus unlikely influenced by broad emission. The widths of the broad components for each mass bin are larger than 200 km s$^{-1}$, thus we consider the shock+precursor models. Additionally, we restrict the models to those that have a pre-shock density $\leq$100 cm$^{-3}$ and to shock velocities within 50 km s$^{-1}$ of the broad component width of a mass bin.

\begin{figure}
    \includegraphics[width=\columnwidth, keepaspectratio]{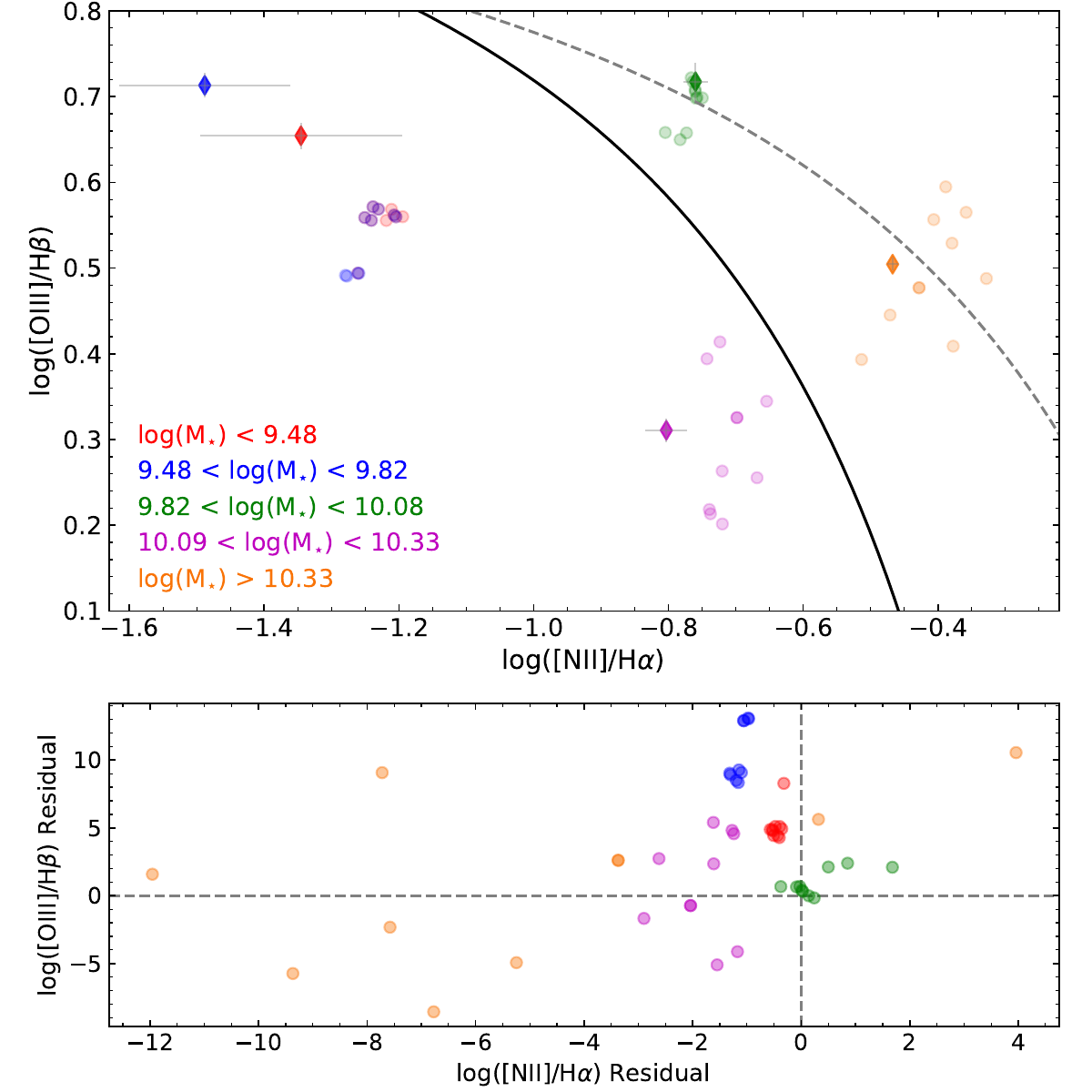}
    \vspace{-0.5cm}
    \caption{BPT diagram for composite spectra binned by stellar mass and shocked+precursor models from the 3MdBs database \protect\citep{Alarie19}. The broad component and ten best shocked+precursor model emission line ratios are shown as diamonds and transparent circles. The solid black line separates star-forming galaxies and AGN \protect\citep{Kauffmann03}. The dashed grey line is the “maximum starburst” line, where above this line lie AGN from \protect\cite{Kewley01}. The bottom panel shows the residuals after subtracting the shocked+precursor models from the observed broad component line ratios.
    }
    \label{fig:BPT}
\end{figure}

To determine whether shocks can explain the observed broad components in the composite spectra, we consider their (1) broad emission line ratios, (2) inferred electron density, and (3) emitting areas. We first compare the measured broad emission line ratios -- \OIII/\HB\ and \NII/\HA\ -- from each mass bin to the predicted shocked+precursor model ratios, keeping the ten models per bin, which best match the observations. Figure \ref{fig:BPT} shows the BPT diagram for the five stellar mass bins and their best-matched models. A majority of the best-matched models for the lowest, second-lowest, and highest mass bins (red, blue, and orange diamonds) are inconsistent with the observed line ratio(s) by $\geq$3$\sigma$. On the other hand, there are models that can reproduce the observed line ratios of the middle and second-highest mass bins (green and purple diamonds) within their uncertainties. In addition to directly comparing line ratios, we can use them to infer other physical properties. If the broad emission arises due to shocks, the electron density measured from the broad [\ion{S}{II}]$\lambda$6717/[\ion{S}{II}]$\lambda$6732 line ratio would come from the post-shock recombination regions where the gas densities are expected to be higher than the pre-shock densities \citep[e.g.,][]{Allen08}. We measure the electron density from the [\ion{S}{II}] ratio of the ten best-matched models for each mass bin using the relationship from \cite{Sanders16}. For the upper three mass bins, the inferred electron densities from the models span over two orders of magnitude, with a majority of the models within 1.5$\sigma$ of $n_{e}$ = 420 cm$^{-3}$ derived in Section \ref{sec:loading}. Conversely, the inferred electron densities in the two lowest mass bins are consistently larger – ranging from $\sim$4000 to 7000 cm$^{-3}$. Finally, the database also provides the predicted \HB\ luminosity per unit area for the different models. Using the best-matched models, we calculate the emitting area required to produce the observed broad component \HB\ luminosity. For the two lowest mass bins, the required emitting areas are $\sim$2 kpc$^{2}$, consistently smaller than the average size of the galaxies in the composite – 8 and 13 kpc$^{2}$, respectively. On the other hand, for the three upper mass bins, the emitting areas are larger, ranging from 4 to 20$\times$ than their average sizes. 

These comparisons show that no shock model simultaneously agrees with the observed broad component line ratios, sample electron density, or the average area of the composite spectra. The best-matched \OIII/\HB\ or \NII/\HA\ model line ratios are inconsistent with the observed broad line ratios in the lowest, second-lowest, and highest mass bins, the inferred model electron densities for the lowest and second-lowest mass bins are $\gtrsim$ 15$\times$ higher than the value we derived in Section \ref{sec:loading}, and the emitting area to reproduce the broad \HB\ luminosity is $\gtrsim$ 4$\times$ the average area of galaxies in the middle, second-highest, and highest mass bins. We therefore conclude that shocks alone cannot produce the broad emission components observed in this study.

This analysis, however, does not rule out other possible origins for the broad emission component. The broad emission may be a mixture of outflowing gas, shocked gas, and other turbulent motions, such as turbulent mixing layers between hot and cold outflow phases. Recent studies that trace outflows using both blueshifted rest-UV absorption lines and broad components of rest-optical emission lines in individual galaxies have found that the two tracers are kinematically similar, thus the broad components can measure the kinematics of outflowing gas \citep{Perrotta21, Avery22}. 

\subsubsection{Other Physical Origins of Scatter}
\label{sec:dyn_range}

In addition to effect of turbulent motions, the lack of correlations between ionised outflow velocity and galactic properties may be due to line-of-sight affects or the small dynamic range probe by the sample \citep[see discussion in][]{Davies19}. If the observations are not well aligned with the outflowing material, then the derived velocities would not capture the true velocities. This in turn would increase the scatter between outflow velocity and galaxy properties and mask any potential correlations. Similarly, if relationships between outflow velocity and properties are weak, then they may not be captured over the small dynamic range probe by our sample. Decomposing rest-optical emission lines into separate narrow and broad components is challenging, with the detection of a broad component requiring high signal-to-noise in the desired line. However, galaxies with strong nebular emission lines are also associated with elevated star-formation properties. Outflows from galaxies with low star-formation properties would be missed if their velocity is low (FWHM$_{\rm{br}}$ < 300 km s$^{-1}$), such that the emission from the broad component is indistinguishable from that of HII regions. As shown in Figure \ref{fig:distribution}, the galaxies with a detected broad component appear to be biased toward higher SFR and \SSFR\ compared to the remaining MOSDEF-ionised galaxies. In comparison, at low redshifts, \cite{Arribas14} measured ionised outflows in luminous and ultra-luminous infrared galaxies and found that outflow velocity scales weakly with both SFR and \SSFR, covering three order-of-magnitude in SFR and \SSFR. Similarly, \cite{Xu22} found the ionised outflow velocity scales weakly with SFR in local low-mass (10$^{4}$ -- 10$^{7}$ $M_{\odot}$) galaxies. In addition to probing larger dynamic ranges, both of these studies detect broad components in lower SFR (\SSFR) galaxies, with a majority below 10 $M_{\odot}$ yr$^{-1}$ (1 $M_{\odot}$ yr$^{-1}$ kpc$^{-2}$). However, the weak scaling relations found by these studies suggest that the outflows are primarily energy-driven, as opposed to the mix of energy- and momentum-driven found for our sample (Section \ref{sec:dirving}). In this case, the scaling between ionised outflow velocity and star-formation properties should be steeper than for a pure energy-driven outflow, thus a relation between ionised outflow velocity and star-formation properties may be captured over a smaller dynamic range. If the outflows are driven by a mix of energy and momentum, then the dynamic range probed by our sample is unlikely the reason why we find no correlation between outflow velocity and star-formation properties.

\section{Conclusions}
\label{sec:conclusions}

We have presented an analysis on the kinematics and energetics of ionised gas outflows within a sample of 598 typical star-forming galaxies at $z$ = 1.4 -- 3.8. Using the extensive spectroscopic dataset of the MOSDEF survey, we decompose strong nebular emission lines of individual galaxies and composite spectra into narrow and broad Gaussian components, tracing virial motions within the galaxy and outflowing gas. Maximum ionised outflow velocities are derived from the FWHM of the broad components, with a mean \Vmax\ = $-$320$\pm$90 km s$^{-1}$. Using \Vmax\ in the outflow model described in \cite{Genzel11}, we estimate the mass-, energy- and momentum-loading factors of the ionised gas outflows. Our main conclusions are as follows:

\begin{description}[leftmargin = 1em]
    \setlength\itemsep{0.75em}
    \item[$\bullet$] There is significant evidence for broad emission components in 10\% (7\%) of \OIII\ (\HA) detections, with their incidence becoming more prevalent among systems with higher SFR and \SSFR. 
    \item[$\bullet$] In individual galaxies, the maximum ionised outflow velocity is not significantly correlated with any galactic property. Composite spectra binned by stellar mass, SFR, and \SSFR\ show clear trends with \Vmax, such that faster outflows are found in bins of higher galactic properties.
    \item[$\bullet$] The ionised outflows appear to remove more momentum than kinetic energy from the ISM, with \Pload/\Eload$\approx$6.5, suggesting that the outflows are primarily momentum-driven. However, these results are model-dependent, and it is likely that there are non-negligible contributions from mechanical energy.
    \item[$\bullet$] The mass-loading factor is marginally correlated (2.5$\sigma$) with stellar mass, scaling as \Mload\ $\propto M_{\star}^{-0.45}$. This scaling is intermediate between the $M_{\star}^{-2/3}$ and $M_{\star}^{-1/3}$ dependence predicted for energy- or momentum-drive outflows, suggesting that these ionised outflows are driven by a combination of these mechanisms, with a larger contribution from radiation pressure acting on cool, dusty material. 
    \item[$\bullet$] We find $\sim$1$\sigma$ agreement between our derived \Mload\ $-$ $M_{\star}$ relation and the theoretical prediction from the FIRE-2 simulations for warm outflowing gas, measured at a larger radial distance.
    \item[$\bullet$] 11\% (0\%) of \OIII\ (\HA) maximum outflow velocities are larger than the escape velocity from the gravitational potential of their host, suggesting that the ionised outflows are often retained and likely setup a galactic fountain. 
    
\end{description}

Obtaining robust constraints on the properties of outflows across different phases is crucial to understand their impact on galaxy evolution. Here, we have studied ionised outflows from typical $z \sim$ 2 galaxies traced by broad components of rest-optical emission lines, finding that these outflows appear to play a negligible role, even during the peak of cosmic star-formation activity. A majority of the ionised outflows likely remain bound to the host galaxy, and their properties (e.g., \Vmax, \Mload) are independent or weakly correlated with galactic properties. However, these results are based upon simple models and assumptions on the geometry and physical condition of the outflowing gas. To build a better understanding of outflows, higher resolution spectroscopic data and spatially resolved imaging is necessary to constrain the geometry, extent, and conditions of outflowing gas. Future progress will greatly benefit from such observations, probing both a wider dynamic range of galaxy properties and tracers of other outflow phases.

\section*{Acknowledgements}

We thank the anonymous referee for providing constructive feedback that improved the paper. The MOSDEF team acknowledges support from NSF AAG grants AST-1312780, 1312547, 1312764, and 1313171, grant AR-13907 from the Space Telescope Science Institute, and grant NNX16AF54G from the NASA ADAP program. We thank the 3D-HST Collaboration, which provided the spectroscopic and photometric catalogs used to select the MOSDEF targets and derive stellar population parameters. This work made use of Astropy:\footnote{\url{http://www.astropy.org}} a community-developed core Python package and an ecosystem of tools and resources for astronomy \citep{Astropy13, Astropy18, Astropy22}. We wish to extend special thanks to those of Hawaiian ancestry on whose sacred mountain we are privileged to be guests. Without their generous hospitality, most of the observations presented herein would not have been possible.

\section*{Data Availability}

In this work, we use 1D spectra and rest-frame optical line measurements from  MOSFIRE Deep Evolution Field (MOSDEF) survey \citep{MOSDEF}. This data is publicly available and can be obtained at \url{http://mosdef.astro.berkeley.edu/for-scientists/data-releases/}.



\bibliographystyle{mnras}
\bibliography{ref} 




\appendix

\section{Derivation of Energy and Momentum Rates}
\label{sec:derivation}

In this section, we derive equations for the energy- and momentum-loading factors of an outflow. A loading factor is the ratio of how much of a quantity (mass, energy, momentum, etc.) is carried out in an outflow relative to the amount produced within a galaxy, thus we define loading factor ($\eta$) as:
\begin{equation}
    \eta_{X} = \frac{\dot{X}_{\rm{out}}}{\dot{X}_{\rm{ref}}}
\end{equation}
where “out” and “ref” refer to the outflow rate and reference rate of the galaxy, respectively.

The energy and momentum outflow rates are easily calculated from the mass outflow rate (Section \ref{sec:loading}) as:
\begin{equation}
    \dot{E}_{\rm{out}} = \frac{1}{2} \dot{M}_{\rm{out}} V_{\rm{out}}^{2}\\
    \dot{p}_{\rm{out}} = \dot{M}_{\rm{out}} V_{\rm{out}}
\end{equation}
On the other hand, estimating the reference energy and momentum rates is non-trivial due to the wide range of physical processes that can generate energy and momentum. Focusing on type IIa supernovae, the total energy injected into the ISM from supernovae is:
\begin{equation}
    \begin{split}
        \dot{E}_{\rm{ref}} = 
        \dot{E}_{\rm{SN}} = \dot{N}_{\rm{SN}}E_{\rm{SN}}
        \approx
        3.17\times10^{41}\left(\frac{\text{SFR}_{\rm{narrow}}}{M_{\odot}\ \text{yr}^{-1}} \right) [\rm{erg}/\rm{s}]
    \end{split}
\end{equation}
where $\dot{N}_{\rm{SN}}$ is the supernova rate, taken as one SN occurs per 100M$_{\odot}$ formed, and $E_{\rm{SN}}$ = 10$^{51}$ ergs is the mechanical energy released by a Type IIa SN.

Following the analytical arguments of \cite{Murray05}, for star-forming galaxies, we consider momentum injection from supernovae and a central starburst:
\begin{equation}
    \label{equ:ploading}
    \begin{split}
        \dot{p}_{\rm{ref}}
        \rightarrow
        \dot{p}_{\rm{SN}} + \dot{p}_{\rm{starburst}}
        \approx
        \sqrt{2E_{\rm{SN}}M_{\rm{ej}}}\dot{N}_{\rm{SN}} + L_{\rm{bol}}/c\\
        \approx
        3.285\times10^{33}\left(\frac{\text{SFR}_{\rm{narrow}}}{M_{\odot}\ \text{yr}^{-1}} \right) [\text{cm g}\ \rm{s}^{-2}]
    \end{split}
\end{equation}
where $M_{\rm{ej}}$ = 10M$_{\odot}$ is the mean ejected mass from a Type II SN, and $L_{\text{bol}}$ is the bolometric luminosity of the galaxy. We assume that $L_{\text{bol}} \sim$ SFR$\times$10$^{10}$ L$_{\odot}$ \citep{Kennicutt98}. 

Alternatively, based on high resolution “local path” simulations of supernovae in the ISM, \cite{Kim15} argue that the $\dot{p}_{\text{sn}}$ is better represented by the spherical momentum at the end of the Sedov-Taylor stage when an SN blast wave cools and a shell forms, rather than the initial injected momentum as in Equation \ref{equ:ploading}. If this convention is adopted then:
\begin{equation}
    \begin{split}
        \dot{p}_{\rm{ref}}
        \rightarrow
        \dot{p}_{\rm{SN}} + \dot{p}_{\rm{starburst}}
        \approx
        \dot{N}_{\rm{SN}}\frac{E_{\rm{SN}}}{v_{\rm{cool}}} + L_{\rm{bol}}/c
        \approx\\
        1.71\times10^{34}\left(\frac{\text{SFR}_{\rm{narrow}}}{M_{\odot}\ \text{yr}^{-1}} \right) [\text{cm g}\ \rm{s}^{-2}]
    \end{split}
\end{equation}
where $v_{\rm{cool}}$ = 200 km s$^{-1}$ is the terminal velocity of the supernova remnant after it has shocked and swept up ambient ISM material.

\section{Results}
\label{sec:tables}

In this Appendix, we provide tables of the ionised gas outflow properties detected in \OIII\ and \HA.

\newcommand{\FA}{\tnotex{tn:1}}
\begin{table*}
  \caption{Properties of the ionised gas based on \OIII\ modelling}
  \label{tbl:OIII}
  \begin{threeparttable}
  \begin{tabular}{lcccccccc}
    \hline
    \hline
    FIELD & V4ID & $z$   & log($M_{\star}/M_{\odot}$) & log(SFR[SED]) & log($\Sigma_{\text{SRF[SED]}}$) & FWHM$_{\text{br}}$ & $V_{\text{off}}$ & $V_{\text{max}}$\\
          &      &     &                            &               &                                 & [km s$^{-1}$]      & [km s$^{-1}$]    & [km s$^{-1}$]\\
    (1)   & (2)  & (3) & (4)                        & (5)           & (6)                             & (7)                & (8)              & (9)\\   
    \hline
    AEGIS   & 3668  & 2.19 &  9.99$\pm$0.04 & 1.38$\pm$0.03 & -0.12$\pm$0.03 & 305.60$_{-4.20}^{+9.17}$     & -9.56$_{-10.33}^{+8.80}$   & -269.11$_{-10.93}^{+11.75}$ \\
    AEGIS   & 4711  & 2.18 &  9.31$\pm$0.04 & 0.76$\pm$0.03 & -0.15$\pm$0.04 & 303.21$_{-2.43}^{+5.34}$     &  26.40$_{-7.14}^{+8.43}$   & -231.12$_{-7.43}^{+9.58}$ \\
    AEGIS   & 6315  & 2.23 &  9.48$\pm$0.04 & 0.87$\pm$0.04 & -0.36$\pm$0.06 & 326.70$_{-17.53}^{+23.51}$   &  85.45$_{-13.54}^{+9.94}$  & -192.02$_{-20.12}^{+22.31}$ \\
    AEGIS   & 10832 & 2.30 &  9.59$\pm$0.05 & 0.98$\pm$0.03 & -0.60$\pm$0.05 & 307.94$_{-5.97}^{+13.57}$    & -75.15$_{-14.03}^{+14.24}$ & -336.70$_{-14.91}^{+18.32}$ \\
    AEGIS   & 12311 & 2.13 &  9.99$\pm$0.04 & 0.87$\pm$0.03 & \ldots\FA      & 325.64$_{-18.84}^{+34.51}$   &  70.63$_{-19.42}^{+18.28}$ & -205.95$_{-25.16}^{+34.54}$ \\
    AEGIS   & 12512 & 2.32 & 10.25$\pm$0.01 & 0.84$\pm$0.01 & -0.87$\pm$0.02 & 424.79$_{-30.55}^{+33.49}$   & -61.81$_{-17.32}^{+16.20}$ & -422.60$_{-31.20}^{+32.73}$ \\
    AEGIS   & 14957 & 2.30 &  9.91$\pm$0.04 & 0.68$\pm$0.03 & -0.83$\pm$0.06 & 338.23$_{-25.36}^{+42.29}$   & -10.28$_{-15.03}^{+14.97}$ & -297.55$_{-26.26}^{+38.91}$ \\
    AEGIS   & 15737 & 2.30 &  9.59$\pm$0.05 & 1.06$\pm$0.03 & -0.11$\pm$0.05 & 302.44$_{-1.83}^{+3.77}$     &  57.84$_{-6.31}^{+7.07}$   & -199.03$_{-6.50 }^{+7.76}$ \\
    AEGIS   & 18543 & 2.14 &  9.35$\pm$0.03 & 0.93$\pm$0.04 &  0.23$\pm$0.05 & 327.0$_{-20.22}^{+48.10}$    &  17.06$_{-12.50}^{+15.20}$ & -260.67$_{-21.24}^{+43.59}$ \\
    AEGIS   & 22931 & 2.30 &  9.41$\pm$0.03 & 1.00$\pm$0.03 &  0.00$\pm$0.05 & 334.46$_{-21.55}^{+31.09}$   & -41.02$_{-11.68}^{+9.32}$  & -325.08$_{-21.71}^{+28.0}$ \\
    AEGIS   & 22935 & 2.37 &  9.71$\pm$0.05 & 0.90$\pm$0.03 & -0.08$\pm$0.06 & 304.97$_{-3.73}^{+8.30}$     &  63.15$_{-9.66}^{+10.96}$  & -195.87$_{-10.16}^{+13.03}$ \\
    AEGIS   & 25817 & 2.29 &  9.68$\pm$0.05 & 0.95$\pm$0.03 & \ldots\FA      & 371.98$_{-41.06}^{+47.02}$   &  29.60$_{-15.98}^{+18.67}$ & -286.33$_{-38.36}^{+44.08}$ \\
    AEGIS   & 30758 & 2.13 &  9.24$\pm$0.05 & 0.63$\pm$0.03 &  0.08$\pm$0.06 & 317.29$_{-13.04}^{+26.89}$   & -16.65$_{-11.66}^{+10.07}$ & -286.14$_{-16.08}^{+24.96}$ \\
    AEGIS   & 35056 & 1.65 &  9.25$\pm$0.05 & 0.50$\pm$0.05 & \ldots\FA      & 314.73$_{-10.59}^{+18.56}$   & -73.53$_{-15.19}^{+14.88}$ & -340.84$_{-17.65}^{+21.68}$ \\
    COSMOS  & 3626  & 2.32 &  9.61$\pm$0.04 & 1.20$\pm$0.02 &  0.72$\pm$0.04 & 563.13$_{-97.07}^{+129.91}$  & -36.57$_{-28.79}^{+26.09}$ & -514.85$_{-87.33}^{+113.38}$ \\
    COSMOS  & 6283  & 2.22 &  9.55$\pm$0.04 & 1.02$\pm$0.03 & -0.11$\pm$0.04 & 333.73$_{-25.29}^{+54.21}$   & -35.38$_{-19.14}^{+12.33}$ & -318.82$_{-28.77}^{+47.66}$ \\
    COSMOS  & 6750  & 2.13 &  9.99$\pm$0.03 & 1.08$\pm$0.02 & -0.67$\pm$0.03 & 555.26$_{-137.63}^{+170.20}$ & -80.12$_{-14.34}^{+22.35}$ & -551.71$_{-117.77}^{+146.27}$ \\
    COSMOS  & 7065  & 3.26 &  9.41$\pm$0.05 & 1.29$\pm$0.03 &  0.28$\pm$0.08 & 325.84$_{-15.56}^{+20.15}$   & -7.90$_{-5.29}^{+4.77}$    & -284.64$_{-14.24}^{+17.77}$ \\
    COSMOS  & 9971  & 2.41 & 10.27$\pm$0.04 & 1.46$\pm$0.02 &  0.23$\pm$0.04 & 327.87$_{-16.37}^{+22.75}$   &  29.92$_{-12.01}^{+15.03}$ & -248.55$_{-18.37}^{+24.48}$ \\
    COSMOS  & 10550 & 3.59 &  9.53$\pm$0.10 & 0.93$\pm$0.04 & -0.14$\pm$0.19 & 487.81$_{-67.77}^{+88.40}$   & -13.33$_{-26.04}^{+20.69}$ & -427.64$_{-63.17}^{+77.88}$ \\
    COSMOS  & 11530 & 2.10 &  9.14$\pm$0.04 & 0.93$\pm$0.04 &  0.25$\pm$0.06 & 391.85$_{-31.46}^{+38.19}$   & -38.60$_{-11.0}^{+8.88}$   & -371.40$_{-28.90}^{+33.63}$ \\
    COSMOS  & 12476 & 1.51 &  9.98$\pm$0.03 & 1.06$\pm$0.04 & -0.31$\pm$0.04 & 397.88$_{-31.78}^{+37.68}$   & -6.93$_{-10.11}^{+9.13}$   & -344.86$_{-28.82}^{+33.28}$ \\
    COSMOS  & 18064 & 1.65 &  9.88$\pm$0.03 & 1.06$\pm$0.03 & -0.25$\pm$0.03 & 431.96$_{-37.31}^{+38.40}$   &  94.97$_{-7.50}^{+3.63}$   & -271.91$_{-32.57}^{+32.82}$ \\
    COSMOS  & 19439 & 2.47 & 10.31$\pm$0.05 & 0.90$\pm$0.02 & \ldots\FA      & 314.28$_{-10.47}^{+20.70}$   & -26.56$_{-8.30}^{+8.20}$   & -293.49$_{-12.17}^{+19.40}$ \\
    COSMOS  & 19985 & 2.19 & 10.32$\pm$0.05 & 1.61$\pm$0.03 &  0.55$\pm$0.03 & 525.70$_{-14.14}^{+14.50}$   & -11.07$_{-4.0}^{+3.68}$    & -457.56$_{-12.66}^{+12.86}$ \\
    COSMOS  & 22576 & 3.26 &  9.38$\pm$0.06 & 1.27$\pm$0.04 &  0.40$\pm$0.06 & 307.98$_{-5.77}^{+10.55}$    &  25.22$_{-4.63}^{+4.81}$   & -236.36$_{-6.74}^{+10.17}$ \\
    COSMOS  & 22838 & 3.36 & 10.24$\pm$0.05 & 1.12$\pm$0.02 &  0.20$\pm$0.07 & 474.56$_{-37.63}^{+45.50}$   & -76.70$_{-14.57}^{+15.30}$ & -479.75$_{-35.12}^{+41.56}$ \\
    COSMOS  & 22862 & 3.12 & 10.01$\pm$0.05 & 1.70$\pm$0.03 & \ldots\FA      & 353.63$_{-28.37}^{+34.33}$   &  71.54$_{-14.0}^{+15.63}$  & -228.81$_{-27.87}^{+33.08}$ \\
    GOODS-N & 1975  & 2.36 &  9.69$\pm$0.05 & 1.17$\pm$0.03 &  0.20$\pm$0.05 & 340.86$_{-29.14}^{+54.59}$   &  39.82$_{-13.81}^{+16.69}$ & -249.68$_{-28.34}^{+49.28}$ \\
    GOODS-N & 16060 & 1.52 & 10.04$\pm$0.06 & 1.12$\pm$0.05 &  0.29$\pm$0.05 & 381.20$_{-32.01}^{+44.41}$   & -1.54$_{-11.78}^{+12.62}$  & -325.30$_{-29.63}^{+39.78}$ \\
    GOODS-N & 22065 & 3.13 &  9.58$\pm$0.06 & 1.17$\pm$0.03 &  0.13$\pm$0.04 & 306.93$_{-5.13}^{+11.35}$    & -41.61$_{-10.22}^{+8.93}$  & -302.29$_{-11.11}^{+13.14}$ \\
    GOODS-N & 22235 & 2.43 &  9.54$\pm$0.05 & 1.12$\pm$0.03 &  0.17$\pm$0.04 & 304.05$_{-3.01}^{+6.11}$     &  59.46$_{-7.72}^{+8.46}$   & -198.78$_{-8.13}^{+9.93}$ \\
    GOODS-N & 24328 & 2.41 &  9.35$\pm$0.05 & 1.24$\pm$0.04 & -0.94$\pm$0.11 & 380.65$_{-40.67}^{+43.42}$   &  33.14$_{-14.98}^{+16.66}$ & -290.16$_{-37.65}^{+40.46}$ \\
    GOODS-N & 27035 & 2.42 &  9.67$\pm$0.03 & 1.06$\pm$0.03 &  0.12$\pm$0.05 & 416.74$_{-27.45}^{+30.97}$   & -92.45$_{-5.44}^{+9.60}$   & -446.40$_{-23.94}^{+28.0}$ \\
    GOODS-N & 30053 & 2.25 & 10.50$\pm$0.03 & 1.08$\pm$0.02 & -0.42$\pm$0.03 & 510.27$_{-47.01}^{+52.94}$   & -10.41$_{-13.81}^{+14.81}$ & -443.80$_{-42.25}^{+47.34}$ \\
    GOODS-N & 34699 & 2.20 & 10.04$\pm$0.06 & 1.43$\pm$0.03 &  0.21$\pm$0.04 & 567.82$_{-90.32}^{+104.20}$  &  4.06$_{-25.61}^{+29.38}$  & -478.21$_{-80.88}^{+93.25}$ \\
    GOODS-N & 35924 & 2.43 &  9.90$\pm$0.05 & 0.98$\pm$0.03 & -1.05$\pm$0.04 & 353.50$_{-22.43}^{+25.66}$   & -31.95$_{-11.01}^{+10.21}$ & -332.18$_{-22.0}^{+24.07}$ \\
    GOODS-S & 40768 & 2.30 & 10.09$\pm$0.02 & 1.58$\pm$0.01 & \ldots\FA      & 302.48$_{-1.85}^{+4.12}$     &  65.98$_{-7.66}^{+8.98}$   & -190.92$_{-7.82}^{+9.64}$ \\
    GOODS-S & 45531 & 2.31 &  9.71$\pm$0.05 & 0.90$\pm$0.02 & -0.24$\pm$0.04 & 352.36$_{-32.24}^{+56.27}$   &  14.60$_{-12.54}^{+12.66}$ &  -284.67$_{-30.12}^{+49.44}$ \\
    \hline
  \end{tabular}
  \begin{tablenotes}
        \item (1): CANDELS field (2): 3D-HST v4 catalogue ID (3): Redshift measured by MOSDEF Survey (4): Stellar Mass (5): Star-formation rate from SED fitting (6): Star-formation-rate surface density (7): Full width half max of the broad \OIII\ component (8): Velocity offset between the broad and narrow \OIII\ components (9): Maximum outflow velocity.
        \item [a] \label{tn:1} Galaxy does not have a robust \RE\ measurement.
    \end{tablenotes}
  \end{threeparttable}
\end{table*}
\begin{landscape}
 \begin{table}
  \caption{Properties of the ionised gas based on \HA\ modelling}
  \label{tbl:HA}
  \begin{threeparttable}
  \begin{tabular}{lcccccccccccc}
    \hline
    \hline
    FIELD & V4ID & $z$   & log($M_{\star}/M_{\odot}$) & log(SFR[H$\alpha$]) & log($\Sigma_{\text{SRF[H$\alpha$]}}$) & FWHM$_{\text{br}}$ & $V_{\text{off}}$ & $V_{\text{max}}$ & $\dot{M}_{\text{out}}$ & log($\eta_{M}$) & log($\eta_{E}$) & log($\eta_{p}$)\\
        &     &     &     &     &     & [km s$^{-1}$]  & [km s$^{-1}$]  & [km s$^{-1}$] & [$M_{\odot}$ yr$^{-1}$] & & & \\
    (1) & (2) & (3) & (4) & (5) & (6) & (7) & (8) & (9) & (10) & (11) & (12) & (13)\\   
    \hline
    AEGIS   & 3658  & 2.17 &  9.86$\pm$0.03 & 1.43$\pm$0.08 & -0.11$\pm$0.08 & 461.09$_{-74.36}^{+95.90}$ & -56.21$_{-26.55}^{+25.92}$ & -447.83$_{-68.50}^{+85.47}$ &  2.53$_{-1.10}^{+1.38}$ & -0.91$_{-0.19}^{+0.27}$ & -1.63$_{-0.26}^{+0.51}$ & -0.98$_{-0.23}^{+0.38}$ \\
    AEGIS   & 8907  & 1.59 & 10.06$\pm$0.03 & 0.87$\pm$0.07 & -1.43$\pm$0.08 & 314.60$_{-11.06}^{+24.10}$ &  17.53$_{-20.49}^{+25.49}$ & -249.68$_{-22.54}^{+32.69}$ &  0.13$_{-0.04}^{+0.04}$ & -1.67$_{-0.13}^{+0.14}$ & -2.88$_{-0.18}^{+0.23}$ & -1.99$_{-0.15}^{+0.18}$ \\
    AEGIS   & 10386 & 1.67 & 10.60$\pm$0.01 & 1.99$\pm$0.05 &  0.05$\pm$0.06 & 392.99$_{-42.97}^{+50.06}$ &  13.52$_{-14.14}^{+17.51}$ & -320.25$_{-39.14}^{+45.98}$ & 10.32$_{-2.94}^{+3.11}$ & -0.65$_{-0.16}^{+0.23}$ & -1.64$_{-0.22}^{+0.35}$ & -0.86$_{-0.19}^{+0.29}$ \\
    AEGIS   & 11930 & 1.57 &  9.70$\pm$0.04 & 1.54$\pm$0.04 & -0.33$\pm$0.05 & 359.54$_{-6.19}^{+6.20}$   & -95.68$_{-3.0}^{+4.42}$    & -401.04$_{-6.05}^{+6.88}$   &  5.58$_{-0.58}^{+0.58}$ & -0.43$_{-0.02}^{+0.02}$ & -1.22$_{-0.03}^{+0.03}$ & -0.54$_{-0.03}^{+0.03}$ \\
    AEGIS   & 14536 & 1.57 & 10.13$\pm$0.06 & 1.51$\pm$0.06 & -0.09$\pm$0.06 & 357.73$_{-14.4}^{+15.25}$  & -19.99$_{-5.02}^{+4.54}$   & -323.82$_{-13.22}^{+13.72}$ &  4.74$_{-0.74}^{+0.75}$ & -0.56$_{-0.06}^{+0.06}$ & -1.54$_{-0.08}^{+0.09}$ & -0.77$_{-0.07}^{+0.07}$ \\
    AEGIS   & 15737 & 2.30 &  9.59$\pm$0.05 & 1.48$\pm$0.05 &  0.30$\pm$0.06 & 327.22$_{-19.68}^{+36.76}$ &  50.49$_{-14.43}^{+18.61}$ & -227.42$_{-22.08}^{+36.35}$ &  3.81$_{-0.99}^{+1.06}$ & -0.70$_{-0.13}^{+0.15}$ & -1.99$_{-0.19}^{+0.27}$ & -1.06$_{-0.16}^{+0.21}$ \\
    AEGIS   & 17556 & 1.40 & 10.13$\pm$0.04 & 1.72$\pm$0.03 & -0.16$\pm$0.03 & 360.59$_{-41.06}^{+65.28}$ &   0.20$_{-13.57}^{+13.64}$ & -306.06$_{-37.42}^{+57.09}$ &  2.63$_{-0.94}^{+1.08}$ & -1.17$_{-0.18}^{+0.23}$ & -2.22$_{-0.24}^{+0.41}$ & -1.41$_{-0.21}^{+0.33}$ \\
    AEGIS   & 26886 & 1.64 & 10.07$\pm$0.04 & 1.26$\pm$0.14 & -0.85$\pm$0.14 & 344.91$_{-28.85}^{+39.11}$ &  77.06$_{-19.96}^{+15.47}$ & -215.88$_{-31.60}^{+36.64}$ &  1.28$_{-0.48}^{+0.50}$ & -0.70$_{-0.13}^{+0.16}$ & -2.04$_{-0.22}^{+0.31}$ & -1.09$_{-0.18}^{+0.24}$ \\
    AEGIS   & 27627 & 1.67 & 10.54$\pm$0.01 & 1.63$\pm$0.10 & -0.43$\pm$0.10 & 437.90$_{-68.67}^{+85.75}$ & -84.99$_{-10.64}^{+16.10}$ & -456.90$_{-59.29}^{+74.59}$ &  4.52$_{-1.66}^{+1.89}$ & -0.72$_{-0.18}^{+0.23}$ & -1.41$_{-0.23}^{+0.42}$ & -0.78$_{-0.20}^{+0.32}$ \\
    AEGIS   & 35764 & 2.37 & 10.47$\pm$0.03 & 1.80$\pm$0.07 &  0.28$\pm$0.08 & 392.37$_{-40.65}^{+48.92}$ & -91.85$_{-6.04}^{+11.25}$  & -425.10$_{-35.05}^{+43.04}$ & 11.97$_{-2.75}^{+2.98}$ & -0.48$_{-0.11}^{+0.13}$ & -1.23$_{-0.16}^{+0.22}$ & -0.57$_{-0.13}^{+0.17}$ \\
    AEGIS   & 39567 & 1.58 & 10.55$\pm$0.03 & 1.82$\pm$0.07 &  0.01$\pm$0.07 & 320.71$_{-14.97}^{+24.19}$ &  51.56$_{-9.48}^{+11.30}$  & -220.82$_{-15.86}^{+23.45}$ &  5.11$_{-1.04}^{+1.05}$ & -0.84$_{-0.08}^{+0.09}$ & -2.15$_{-0.14}^{+0.17}$ & -1.21$_{-0.11}^{+0.13}$ \\
    COSMOS  & 3185  & 2.17 &  9.74$\pm$0.06 & 1.01$\pm$0.26 &  0.50$\pm$0.28 & 397.59$_{-69.38}^{+117.34}$&  64.47$_{-27.25}^{+23.33}$ & -273.22$_{-64.92}^{+102.35}$&  3.03$_{-2.36}^{+3.43}$ & -0.25$_{-0.23}^{+0.34}$ & -1.39$_{-0.33}^{+0.80}$ & -0.53$_{-0.28}^{+0.54}$ \\
    COSMOS  & 6754  & 2.12 &  9.55$\pm$0.04 & 0.33$\pm$0.24 & -1.12$\pm$0.25 & 329.68$_{-21.28}^{+37.72}$ &  71.96$_{-21.83}^{+18.0}$  & -208.05$_{-28.34}^{+36.75}$ &  0.30$_{-0.18}^{+0.19}$ & -0.41$_{-0.12}^{+0.15}$ & -1.77$_{-0.21}^{+0.29}$ & -0.80$_{-0.17}^{+0.22}$ \\
    COSMOS  & 11530 & 2.10 &  9.14$\pm$0.04 & 1.12$\pm$0.10 &  0.44$\pm$0.10 & 337.50$_{-26.18}^{+46.48}$ & -77.23$_{-16.04}^{+23.74}$ & -363.88$_{-27.42}^{+46.06}$ &  2.78$_{-1.03}^{+1.16}$ & -0.56$_{-0.16}^{+0.18}$ & -1.45$_{-0.19}^{+0.28}$ & -0.72$_{-0.17}^{+0.23}$ \\
    COSMOS  & 12476 & 1.51 &  9.98$\pm$0.03 & 1.61$\pm$0.08 &  0.24$\pm$0.08 & 449.34$_{-36.29}^{+47.37}$ & -14.34$_{-8.55}^{+8.65}$   & -395.98$_{-31.98}^{+41.16}$ &  9.99$_{-2.49}^{+2.60}$ & -0.30$_{-0.12}^{+0.14}$ & -1.11$_{-0.16}^{+0.23}$ & -0.42$_{-0.14}^{+0.18}$ \\
    COSMOS  & 13701 & 2.17 & 10.65$\pm$0.05 & 1.98$\pm$0.09 & -0.04$\pm$0.09 & 308.53$_{-6.5}^{+15.82}$   & -65.10$_{-19.91}^{+17.09}$ & -327.14$_{-20.66}^{+21.74}$ &  4.80$_{-1.58}^{+1.82}$ & -1.14$_{-0.16}^{+0.17}$ & -2.12$_{-0.17}^{+0.20}$ & -1.35$_{-0.16}^{+0.18}$ \\
    COSMOS  & 18064 & 1.65 &  9.88$\pm$0.03 & 1.49$\pm$0.09 &  0.17$\pm$0.09 & 420.22$_{-28.61}^{+33.58}$ &  82.91$_{-13.07}^{+11.18}$ & -274.01$_{-27.59}^{+30.64}$ &  5.20$_{-1.47}^{+1.54}$ & -0.48$_{-0.12}^{+0.14}$ & -1.61$_{-0.17}^{+0.24}$ & -0.77$_{-0.14}^{+0.19}$ \\
    COSMOS  & 27945 & 2.02 & 10.10$\pm$0.03 & \ldots\tnotex{tn:a}  & \ldots  & 447.48$_{-31.03}^{+28.44}$ & -97.20$_{-2.09}^{+5.51}$   & -477.26$_{-26.44}^{+24.78}$ & \ldots\tnotex{tn:c}     & \ldots\tnotex{tn:c}     & \ldots\tnotex{tn:c}     & \ldots\tnotex{tn:c}     \\
    GOODS-N & 328   & 2.27 & 10.53$\pm$0.05 & 2.08$\pm$0.04 &  0.47$\pm$0.04 & 322.53$_{-17.79}^{+66.75}$ & -83.99$_{-11.25}^{+17.43}$ & -357.92$_{-18.84}^{+59.31}$ &  5.24$_{-1.91}^{+2.19}$ & -1.30$_{-0.17}^{+0.21}$ & -2.20$_{-0.21}^{+0.32}$ & -1.47$_{-0.19}^{+0.26}$ \\
    GOODS-N & 7652  & 2.27 & 10.59$\pm$0.04 & 1.51$\pm$0.20 & -0.54$\pm$0.20 & 315.10$_{-11.22}^{+25.55}$ &  64.12$_{-14.55}^{+14.21}$ & -203.50$_{-17.39}^{+25.94}$ &  2.00$_{-1.08}^{+1.17}$ & -0.85$_{-0.13}^{+0.16}$ & -2.24$_{-0.18}^{+0.24}$ & -1.26$_{-0.15}^{+0.20}$ \\
    GOODS-N & 8099  & 1.49 & 10.39$\pm$0.01 & 1.78$\pm$0.04 &  0.36$\pm$0.04 & 377.52$_{-29.5}^{+33.72}$  & -21.98$_{-12.56}^{+11.24}$ & -342.62$_{-28.03}^{+30.77}$ &  6.79$_{-1.52}^{+1.62}$ & -0.80$_{-0.11}^{+0.13}$ & -1.74$_{-0.15}^{+0.21}$ & -0.99$_{-0.13}^{+0.16}$ \\
    GOODS-N & 12345 & 2.27 & 10.34$\pm$0.05 & 1.82$\pm$0.11 & -0.13$\pm$0.11 & 331.89$_{-23.18}^{+43.44}$ & -3.62$_{-12.06}^{+12.67}$  & -285.50$_{-23.09}^{+39.01}$ &  3.52$_{-1.23}^{+1.37}$ & -1.11$_{-0.13}^{+0.15}$ & -2.21$_{-0.18}^{+0.27}$ & -1.37$_{-0.15}^{+0.20}$ \\
    GOODS-N & 16060 & 1.52 & 10.04$\pm$0.06 & 1.54$\pm$0.05 &  0.71$\pm$0.06 & 367.17$_{-27.57}^{+45.56}$ & -1.86$_{-9.03}^{+7.95}$    & -313.71$_{-25.10}^{+39.51}$ & 13.52$_{-3.56}^{+3.59}$ & -0.07$_{-0.16}^{+0.21}$ & -1.09$_{-0.20}^{+0.30}$ & -0.29$_{-0.18}^{+0.25}$ \\
    GOODS-N & 22235 & 2.43 &  9.54$\pm$0.05 & 1.23$\pm$0.09 &  0.28$\pm$0.09 & 313.61$_{-9.96}^{+20.69}$  &  72.38$_{-15.0}^{+14.80}$  & -193.97$_{-17.22}^{+22.97}$ &  4.18$_{-1.04}^{+1.06}$ & -0.17$_{-0.10}^{+0.13}$ & -1.59$_{-0.15}^{+0.21}$ & -0.59$_{-0.12}^{+0.16}$ \\
    GOODS-N & 23869 & 2.24 & 10.32$\pm$0.02 & 1.80$\pm$0.16 & -0.08$\pm$0.16 & 317.01$_{-11.14}^{+15.23}$ & -85.73$_{-9.65}^{+12.80}$  & -354.98$_{-13.52}^{+18.19}$ &  8.58$_{-3.54}^{+3.53}$ & -0.51$_{-0.08}^{+0.09}$ & -1.41$_{-0.09}^{+0.11}$ & -0.68$_{-0.08}^{+0.10}$ \\
    GOODS-N & 27035 & 2.42 &  9.67$\pm$0.03 & 1.45$\pm$0.09 &  0.50$\pm$0.10 & 474.07$_{-50.36}^{+58.06}$ & -92.57$_{-5.47}^{+10.62}$  & -495.20$_{-43.12}^{+50.44}$ & 16.43$_{-4.27}^{+4.28}$ &  0.16$_{-0.11}^{+0.14}$ & -0.46$_{-0.16}^{+0.24}$ &  0.13$_{-0.14}^{+0.19}$ \\
    GOODS-N & 28061 & 2.20 &  9.98$\pm$0.03 & 2.32$\pm$0.20 &  0.45$\pm$0.21 & 317.56$_{-13.19}^{+28.75}$ & -47.24$_{-20.20}^{+14.38}$ & -316.96$_{-23.10}^{+28.34}$ & 15.37$_{-8.94}^{+10.42}$& -0.93$_{-0.14}^{+0.17}$ & -1.93$_{-0.17}^{+0.22}$ & -1.14$_{-0.16}^{+0.19}$ \\
    GOODS-N & 30053 & 2.25 & 10.50$\pm$0.03 & 1.86$\pm$0.07 &  0.35$\pm$0.07 & 452.85$_{-32.66}^{+37.16}$ & -29.71$_{-14.87}^{+13.36}$ & -414.32$_{-31.47}^{+34.27}$ & 21.43$_{-4.35}^{+4.18}$ & -0.04$_{-0.10}^{+0.12}$ & -0.81$_{-0.14}^{+0.18}$ & -0.14$_{-0.12}^{+0.15}$ \\
    GOODS-N & 30564 & 2.48 & 10.47$\pm$0.01 & 1.62$\pm$0.07 & \ldots\tnotex{tn:b}   & 382.87$_{-43.9}^{+53.60}$  &  72.10$_{-20.24}^{+17.68}$ & -253.08$_{-42.42}^{+48.84}$ & \ldots\tnotex{tn:c}     & \ldots\tnotex{tn:c}     & \ldots\tnotex{tn:c}     & \ldots\tnotex{tn:c} \\
    GOODS-N & 34699 & 2.20 & 10.04$\pm$0.06 & \ldots\tnotex{tn:a}  & \ldots  & 382.97$_{-36.98}^{+43.54}$ &  25.95$_{-10.21}^{+11.46}$ & -299.32$_{-33.03}^{+38.72}$ & \ldots\tnotex{tn:c}     & \ldots\tnotex{tn:c}     & \ldots\tnotex{tn:c}     & \ldots\tnotex{tn:c} \\
    GOODS-S & 40768 & 2.30 & 10.09$\pm$0.02 & 2.15$\pm$0.03 & \ldots\tnotex{tn:b}   & 320.96$_{-15.34}^{+30.67}$ &  38.31$_{-9.46}^{+12.31}$  & -234.29$_{-16.10}^{+28.81}$ & \ldots\tnotex{tn:c}     & \ldots\tnotex{tn:c}     & \ldots\tnotex{tn:c}     & \ldots\tnotex{tn:c} \\
    GOODS-S & 46938 & 2.33 &  9.95$\pm$0.04 & 1.95$\pm$0.07 & \ldots\tnotex{tn:b}   & 446.84$_{-49.03}^{+58.20}$ &   6.37$_{-16.79}^{+16.21}$ & -373.14$_{-44.91}^{+52.02}$ & \ldots\tnotex{tn:c}     & \ldots\tnotex{tn:c}     & \ldots\tnotex{tn:c}     & \ldots\tnotex{tn:c} \\
    UDS     & 16873 & 1.47 & 10.71$\pm$0.01 & \ldots\tnotex{tn:a}  & \ldots  & 304.89$_{-3.68}^{+7.96}$   & -83.58$_{-10.02}^{+11.19}$ & -342.52$_{-10.50}^{+13.07}$ & \ldots\tnotex{tn:c}     & \ldots\tnotex{tn:c}     & \ldots\tnotex{tn:c}     & \ldots\tnotex{tn:c} \\
    \hline\\
  \end{tabular}
  \begin{tablenotes}
        \item (1): CANDELS field (2): 3D-HST v4 catalogue ID (3): Redshift measured by MOSDEF Survey (4): Stellar Mass (5): Star-formation rate from SED fitting (6): Star-formation-rate surface density (7): Full width half max of the broad \HA\ component (8): Velocity offset between the broad and narrow \HA\ components (9): Maximum outflow velocity (10): Mass outflow rate (11): Mass loading factor (12): Energy loading factor (13): Momentum loading factor
        \item[a] \label{tn:a} Galaxy does not have a significant detection of H$\beta$
        \item[b] \label{tn:b} Galaxy does not have a robust \RE\ measurement
        \item[c] \label{tn:c} Property could not be measured, due to lack of SFR or \RE.
  \end{tablenotes}
  \end{threeparttable}
\end{table}

\end{landscape}


\bsp	
\label{lastpage}
\end{document}